# Bernardo Dessau, fisico, scienziato, maestro, da Bologna a Perugia tra i marosi del secolo breve

Giovanni Carlotti

*Dipartimento di Fisica e Geologia, Università degli Studi di Perugia e CNR-IOM – Istituto Officina dei Materiali, Via A. Pascoli, 06123 Perugia, Italia - giovanni.carlotti@unipg.it*

**Riassunto.** Settantacinque anni fa moriva Bernardo Dessau, che fu per un trentennio l'unico professore di Fisica dell'Università di Perugia. Con questo breve contributo intendiamo saldare un debito di riconoscenza, anzitutto perché i suoi meriti scientifici furono verosimilmente sottovalutati, avendo egli trascorso i primi quindici anni di carriera all'ombra del grande Augusto Righi a Bologna, dove fu anche testimone degli incontri con il giovane Marconi. Egli fu poi relegato presso la Facoltà di Medicina di Perugia, sprovvisto di un vero laboratorio e poco valorizzato dalla comunità scientifica nazionale, senza più riuscire a trasferirsi in sedi meglio attrezzate. Si rivelò comunque un fine didatta e divulgatore, pubblicando un poderoso manuale di fisica in tre volumi che si diffuse in Italia e all'estero. Pur essendo cittadino italiano fin dal 1894, la sua origine tedesca fu causa di sospensione dal servizio durante la prima guerra mondiale, mentre la sua identità ebraica lo espose alle conseguenze dell'antisemitismo nazifascista: fu radiato dall'Università di Perugia e dalla Società Italiana di Fisica, di cui era socio fin dai primordi, scampando fortunosamente alla deportazione.

**Abstract.** Seventy-five years ago, Bernardo Dessau passed away. He had been the sole professor of Physics at the University of Perugia for thirty years. With this brief contribution we intend to repay a debt of gratitude, first of all because his scientific merits were probably underestimated, having spent the first fifteen years of his career under the shadow of the great Augusto Righi in Bologna, where he also witnessed meetings with the young Marconi. He was then relegated to the Faculty of Medicine of Perugia, without a real laboratory, disregarded by the national scientific community and unable to move to better equipped locations. Nevertheless, he proved to be a gifted teacher and popularizer of science. He published a comprehensive three-volume physics textbook that was widely used in Italy and abroad. Despite becoming an Italian citizen in 1894, Dessau's German origins led to his suspension from service during World War I. Furthermore, his Jewish identity made him a target of Nazi-fascist antisemitism. He was dismissed from both the University of Perugia and from the Italian Society of Physics, of which he had been a member since its beginning, fortunately escaping deportation.

**1. Educazione in una famiglia rabbinica e studi di fisica a Berlino e Strasburgo**

Bernardo Dessau viene alla luce il 13 agosto 1863 da una coppia relativamente importante: Samuel Dessau (1826-1904), discendente da una famiglia di rabbini di Amburgo e Fanny Schwarzschild (1832-1893) della nota famiglia di banchieri di Francoforte [1]. Samuel diventa rabbino a Offenbach am Main, nei pressi di Francoforte, ma poi si trasferisce con tutta la famiglia a Fürth, alla periferia di Norimberga, dapprima come professore di scienze naturali alla Realschule ebraica e infine come direttore della Israelitische Bürgerschule.

Bernardo, quinto di sette figli, nasce dunque a Offenbach da questa coppia con profonde radici ebraiche mitteleuropee, ma gran parte della prima giovinezza la trascorre appunto a Fürth, dove l'interesse per gli studi scientifici è favorito dall'attività di insegnamento di suo padre che si occupa soprattutto di chimica. Egli viene avviato alla conoscenza della Bibbia ed agli studi classici, imparando a memoria numerosi brani della sacra scrittura in lingua ebraica, così come brani in greco dell'Iliade e dell'Odissea, che lo accompagneranno lungo tutta la sua esistenza. Anche perché rimarrà fino alla fine scrupoloso osservante delle tradizioni religiose ebraiche, dal rigoroso rispetto del riposo sabbatico a quello delle prescrizioni liturgiche e alimentari previste dalla Tōrāh. A 13 anni si ammala di scarlattina ed è protagonista di un grave incidente: la caduta da un tetto, dove era salito per riprendere un pallone, che gli provoca un blocco renale e lo costringe a una lunga convalescenza. Da allora nella nidiata dei fratelli, diventa il «Sorgenkind» (bambino difficile) della famiglia, per cui la mamma, che ne ha intuito da subito le rare doti di cuore e di ingegno, dedica a lui speciale attenzione e cure particolari ([1]). In effetti, la lunga esistenza di Bernardo sarà costantemente segnata da problemi di salute ricorrenti e persistenti, soprattutto legati a malattie polmonari, mal di stomaco, reumatismi. Tra i fratelli, colui che diventa più famoso di tutti è senza dubbio il primogenito Hermann (1856-1931), che, una volta completati gli studi classici, raggiunge notorietà di epigrafista a livello internazionale come allievo prediletto e continuatore dell'opera di Theodor Mommsen. Bernardo invece, che rivela attitudini tecniche, in un primo momento viene indirizzato dal padre agli studi industriali, ma poi, desideroso di una formazione universitaria, si prepara da solo a sostenere l'esame di maturità classica (Abitur) che supera brillantemente nel 1882 ([2]), iscrivendosi infine alla facoltà di Fisica di Berlino dove è allievo di Herman von Helmholtz. Si trasferisce quindi a Strasburgo e si laurea con August Kundt nel 1886, con un lavoro di tesi che viene pubblicato sugli Annalen der Physik un Chemie, a sola sua firma, sotto il titolo: *Über Metallschichten, welche durch Zerstäuben einer Kathode entsehen* (*Sulla metallizzazione che si ottiene con la disgregazione di un*

---

([1]) Queste notizie, così come altre di carattere privato e familiare che saranno riportate nel corso dell'articolo, mi sono state comunicate direttamente dall'Ing. Umberto Steindler (classe 1936, figlio di Fanny Dessau Steindler, primogenita di Bernardo Dessau), residente a Parigi, durante piacevoli conversazioni in videochiamata, avute nel corso dell'anno 2023, di cui gli sono veramente grato.

([2]) Sempre dai racconti a me riportati direttamente dal nipote Ing. Umberto Steindler, risulta che alla festa per la Abitur di Bernardo, tutti i professori tennero magnifici e dotti discorsi su Socrate, Platone, Goethe e su altri campioni della cultura classica, come usuale, ma ciò che rimase negativamente impresso nella mente del giovane Bernardo fu il nazionalismo fanatico che traspari da parte di coloro che erano stati i suoi maestri, al momento della chiusura della festa quando intonarono l'inno "Deutschland über Alles".

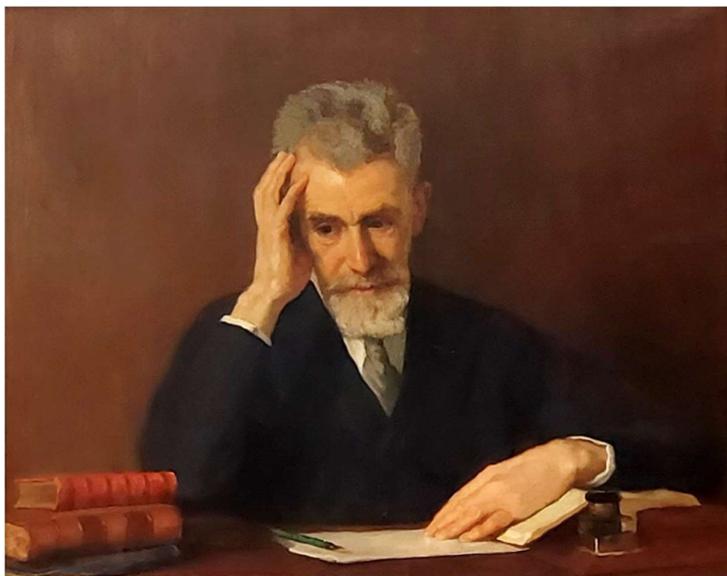

**Fig. 1**. Ritratto di Bernardo Dessau dipinto nel 1936 dalla moglie Emma Goitein Dessau (collezione del Dipartimento di Fisica e Astronomia "Augusto Righi" di Bologna).

*catodo*) [2] ($^3$). Si tratta di un lavoro pionieristico su quella che oggi chiameremmo la tecnica della polverizzazione catodica, o sputtering, che è tuttora ampiamente utilizzata, sia in applicazioni di ricerca che nell'industria, per produrre film sottili e multistrati ($^4$). Nel lavoro di tesi di Dessau, egli, sotto la supervisione di Kundt, realizza specchi metallici depositando, a bassa pressione, strati metallici di oro, argento, platino, rame, nichel o ferro su substrati di vetro, controllandone l'ossidazione e analizzandone alcune caratteristiche fisiche, come la birifrangenza ottica. L'anno seguente, Bernardo si ammala gravemente ai polmoni e per ristabilirsi viene mandato in montagna, a Merano, dove già allora c'erano alberghi con ristorante kasher, cioè in grado di offrire cibi e bevande preparati in base alle leggi alimentari ebraiche. Entra così in contatto anche con la cultura italiana e, una volta ristabilito dalla malattia, su parere dei medici che sconsigliano il clima rigido e umido della Germania per la sua salute, decide di trasferirsi in Italia. Sceglie Pisa, contando da un lato sul clima mite e temperato assicurato dalla vicinanza del mare e dall'altro sulla tradizione culturale e scientifica della città che era stata la patria di Galileo Galilei. Ma l'idea di lasciare la Germania si deve probabilmente anche al disagio avvertito dal giovane Bernardo difronte al montante nazionalismo tedesco e allo strisciante antisemitismo della Germania guglielmina che rendeva quasi impossibile, per un giovane ebreo, adire alla carriera accademica. Basti

---

($^3$) Per la descrizione bibliografica delle pubblicazioni scientifiche scritte da B. Dessau e ricordate all'interno di questo lavoro, si è fatto riferimento al capitolo "La bibliografia di Bernardo Dessau" in [1], p 41

($^4$) Lo sviluppo di questa tecnica trasse vantaggio dal fatto che proprio nella seconda metà del diciannovesimo secolo vi fu un sostanziale e rapido progresso nelle tecniche per ottenere il vuoto. Grazie a tali progressi si poté quindi avviare la deposizione di film sottili in atmosfera pulita, così come altri tipi di esperimenti, quali ad esempio lo studio controllato delle scariche elettriche nei gas rarefatti e lo studio dei raggi catodici, fino alla scoperta dell'elettrone nel 1897.

pensare che il fratello maggiore Hermann, la cui fama di epigrafista abbiamo ricordato in precedenza, rimase libero docente (Privatdozent) fino al 1917 quando, ormai sessantunenne, divenne professore onorario a Berlino, ma non poté mai ottenere la cattedra da professore ordinario proprio a causa delle sue radici ebraiche. A partire dal 1888, Bernardo inizia quindi a frequentare l'ambiente universitario pisano e subito si diffonde la fama della sua preparazione tecnico/scientifica, soprattutto della speciale abilità nel soffiare il vetro per realizzare tubi da vuoto. Proprio in quel periodo, il bolognese Augusto Righi, che nel 1885 si è trasferito a Padova da Palermo, dove ha iniziato la sua carriera come professore ordinario di fisica sperimentale nel 1880, sta cercando un assistente e, venuto a sapere di questo brillante giovanotto tedesco con una solida formazione alle spalle, lo prende con sé nel 1889 presso l'ateneo patavino. L'anno seguente, Righi riesce finalmente a tornare nella nativa Bologna e porta con sé presso l'Alma Mater il giovane Dessau in veste di suo assistente, come risulta dalla nomina ministeriale riprodotta in Fig. 2. Inizia così, nel novembre del 1890, il quindicennio bolognese di Dessau alla "corte" di colui che rapidamente e meritatamente diventerà il fisico sperimentale più famoso d'Italia, con forti legami e ampia notorietà anche sul piano internazionale, al punto che collezionerà ben quindici candidature di fila al premio Nobel, dal 1905 al 1920, anno della sua morte [3].

## 2    Il periodo bolognese alla corte del grande Augusto Righi

### 2.1 I primi cinque anni: incerto se tornare in Germania o trasferirsi stabilmente in Italia

Quando nel 1889 Dessau incontra Righi, quest'ultimo ha quasi quarant'anni e ha alle spalle un'intensa attività di ricerca sperimentale in molti campi emergenti della fisica ottocentesca, dalle proprietà magnetiche dei materiali alle tecniche ottiche, dall'effetto fotoelettrico ai raggi x [4]. Su entrambe quest'ultimi fenomeni, il Righi potrebbe addirittura vantare una sorta di pari-merito nella scoperta (lo stesso termine "effetto fotoelettrico" è suo), malgrado la storia non glieli abbia riconosciuti, come ben ricostruito a p. 215 e p. 403 di Ref. [4]. A partire dal 1890, tuttavia, proprio nel trasferirsi definitivamente a Bologna, dove si era svolta la sua formazione universitaria, Righi volge il suo interesse verso il nascente campo di indagine delle onde elettromagnetiche, trascinando dunque il giovane Bernardo in questa avventura. È noto che l'iniziatore di questo campo di ricerca era stato il giovane fisico tedesco Heinrich Hertz, che nel 1887 era riuscito a generare e ricevere il primo segnale elettromagnetico. Qualche anno dopo, però, egli muore poco più che trentenne, ignaro delle straordinarie conseguenze e applicazioni delle sue scoperte. Righi decide dunque di dedicarsi, nel progettare il proprio lavoro di ricerca a Bologna, all'indagine di queste nuove onde ("oscillazioni elettriche", come venivano chiamate all'epoca) e si pone l'obiettivo di verificare che valgono anche per esse le leggi dell'ottica. A tal fine, capisce che è importante riuscire a generare e rivelare onde di lunghezza d'onda ben più corta di quelle di Hertz, in modo da poter rendere trascurabili gli effetti della diffrazione e poter utilizzare strumenti di dimensioni compatibili con il laboratorio. Grazie alle sue straordinarie capacità sperimentali, nel giro di pochi mesi egli riesce in effetti a sviluppare la tecnica di generazione e rivelazione dei segnali elettromagnetici, riducendo la lunghezza d'onda dai circa 65 cm ottenuti da Hertz a circa 2 cm (Ref. [4], p. 38). In questo modo, porta a compimento esperimenti che dimostrano il rispetto delle leggi della riflessione e della rifrazione, oltre a sperimentare che i segnali si possono propagare fino alla distanza circa 25 m (corrispondente verosimilmente alla massima lunghezza del corridoio dell'istituto). Alcuni degli strumenti originali utilizzati per questi esperimenti sono ben visibili presso il Dipartimento di Fisica e Astronomia (DIFA) di Bologna, dedicato proprio al Righi, dove è stata allestita una ricca e documentata esposizione nel 2020, in occasione del centenario della morte dello scienziato. Fanno dunque bella mostra di sé numerosi apparecchi dell'epoca, dal generatore a scintille allo specchio parabolico utilizzato per concentrare la potenza elettromagnetica sul rivelatore. In questo lavoro certosino e sistematico di verifica delle leggi che regolano la generazione e la propagazione delle onde elettromagnetiche è verosimile che la presenza del giovane Dessau sia stata importante per almeno due ordini di motivi. Anzitutto, abbiamo già accennato alla sua solida esperienza e formazione sperimentale, inclusa l'abilità tecnica nel realizzare tubi di vetro sottovuoto. In secondo luogo, Dessau è in quel periodo il collaboratore principale di Righi, anche come ponte ideale tra la letteratura scientifica germanica, a partire dai pionieristici risultati di Hertz, e le ricerche del suo mentore. Egli, infatti, oltre a svolgere un'intensa attività di laboratorio, in questi primi anni di permanenza a Bologna lavora incessantemente come traduttore in entrambe le direzioni. Da un lato, traduce in italiano per Righi, che non conosce la lingua teutonica, gli articoli più interessanti su questi argomenti che appaiono nella letteratura scientifica tedesca. Ancora oggi, nel museo del DIFA sono conservati alcuni quaderni, come quelli in Fig. 3, dove è riportata la traduzione italiana di articoli apparsi in riviste tedesche. Dall'altro lato, Dessau fa anche il lavoro opposto, di tradurre, cioè, in tedesco i risultati e le pubblicazioni del Righi. Basti pensare al famoso trattato sull'Ottica delle Oscillazioni Elettriche, che Righi pubblica, a sola sua firma, nel 1897 [5] e che subito dopo appare anche nell'edizione tedesca tradotta a opera di Dessau [6]. In Fig. 4 è riprodotta la foto di un'aula universitaria dove Righi e Dessau sono alle prese con un apparato per la generazione delle onde elettromagnetiche.

Del lungo periodo di Dessau a fianco di Righi vanno comunque, a mio giudizio, distinti i primi cinque anni (1889-1894), durante i quali il giovane assistente si mette alla prova ma non ha ancora chiaro quale tipo di carriera perseguire, incerto se rimanere in Italia o provare a tornare nella nativa Germania. Un momento di svolta è rappresentato appunto dall'estate del 1894, quando egli si trova in vacanza a Fürth, presso la propria famiglia, e intesse una fitta corrispondenza con Righi [7]. Nella missiva del 3 agosto Dessau gli comunica che sta vagliando la proposta di un incarico all'Istituto Tecnico di Augsburg, ma lo prega di lasciargli aperta, fino a fine mese, la possibilità di rientrare a Bologna. Tuttavia, già nel giro di una settimana si rende conto che le caratteristiche dell'incarico proposto dal Ministero tedesco della Pubblica Istruzione non corrispondono alle sue aspettative, così nelle lettere del 10 e del 15 agosto chiede con una certa apprensione conferma a Righi di poter eventualmente essere riaccolto a Bologna in vista del nuovo anno accademico. Il giorno dopo, 16 agosto, avendo finalmente ricevuto da Righi conferma della possibile proroga della posizione bolognese, Dessau rompe gli indugi e conferma al suo mentore di poter ancora lavorare per lui per un anno, durante il quale spera di poter conseguire la libera docenza. Tanto per rendere sinteticamente il tipo di rapporto che egli aveva con il suo maestro, basti notare che tutte le lettere si concludono con l'espressione "suo devoto servitore".

**2.2 Dal 1895 al 1904: la scelta di restare in Italia e la faticosa ricerca di una sistemazione lavorativa**

Una volta abbandonata l'opzione di tornare in Germania, l'anno accademico 1894/95 riserva una serie di novità positive per Dessau, tutte documentate nel fascicolo personale presente presso l'archivio storico dell'Università di Bologna. Anzitutto, il 14 ottobre 1894 riceve dall'Ufficio di Stato Civile del Comune di Bologna la comunicazione di aver conseguito la cittadinanza italiana. Nel frattempo, fa passi avanti la procedura avviata a inizio anno con la domanda al Ministero per ottenere la libera docenza in fisica sperimentale. La commissione all'uopo designata in aprile, composta da cinque professori universitari, Ricci, Chistoni, Saporetti, Ciamician e lo stesso Righi, si consulta per la prima volta nel luglio 1894 e indica il seguente argomento per la dissertazione: *"La teoria elettromagnetica della luce, almeno in quanto ai concetti fondamentali ed alle conferme date dall'esperienza"*.

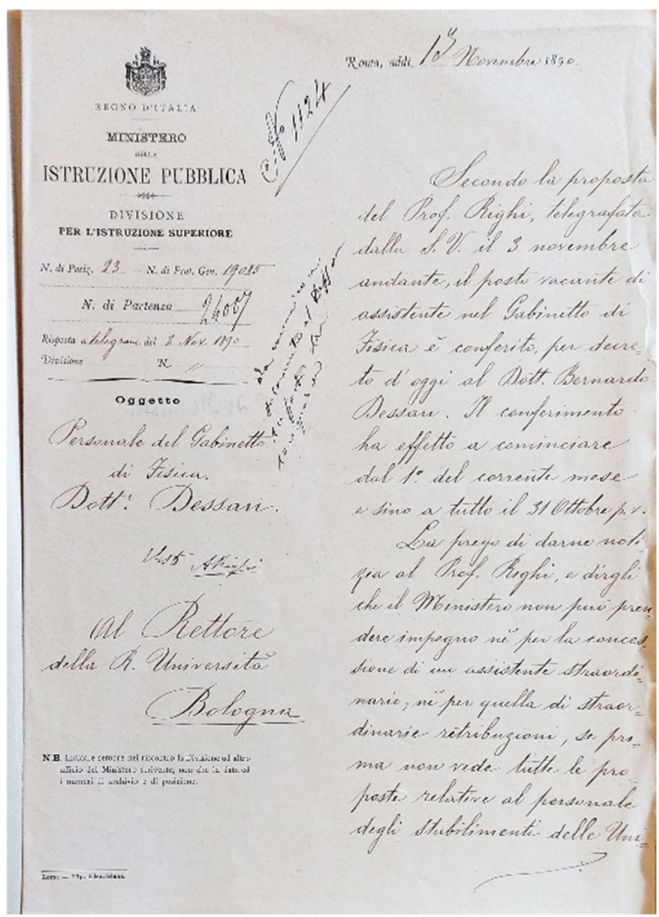

**Fig. 2** Nomina ministeriale di Bernardo Dessau come assistente del Gabinetto di Fisica di Bologna, datata 13 novembre 1890. Su concessione della Alma Mater Studiorum Università di Bologna – Archivio storico. Divieto di ulteriore riproduzione o duplicazione con qualsiasi mezzo.
.

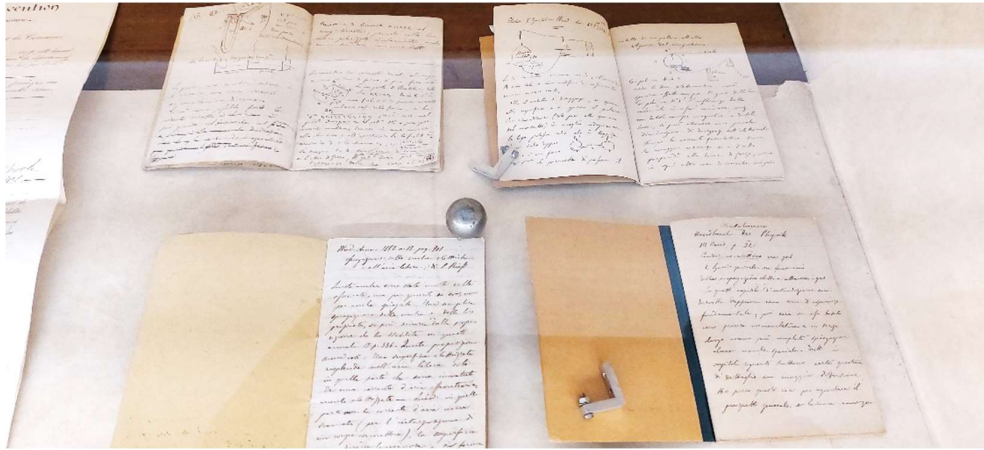

**Fig.3** Foto di alcuni quaderni contenenti la traduzione in italiano di articoli scientifici apparsi su riviste tedesche, tradotti da Dessau, come si riconosce dalla calligrafia (collezione del Dipartimento di Fisica e Astronomia "Augusto Righi" di Bologna).

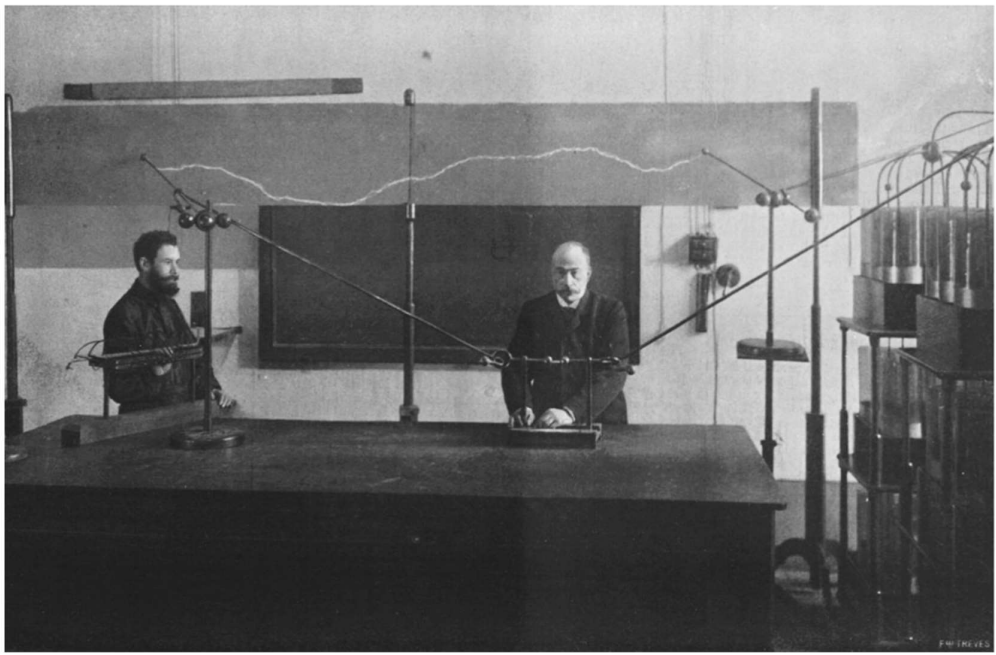

**Fig.4** Fotografia di Augusto Righi (al centro) e Bernardo Dessau (a sinistra), alle prese con l'apparato per la generazione delle onde elettromagnetiche. ([8], p. 171)

Sebbene da regolamento fosse sufficiente attendere un periodo di tre mesi prima di procedere con la discussione e l'esame, bisognerà attendere l'estate successiva per la discussione della tesi di abilitazione (9 e 10 luglio 1895), che la commissione approva con la votazione di 45/50. Dopo ulteriori tre mesi, in ottobre, il Rettore di Bologna trasmette i verbali al Ministero, il quale però a novembre dello stesso anno ravvisa un'irregolarità nei verbali e chiede alla commissione di chiarire alcuni aspetti della procedura, per cui bisognerà attendere addirittura il 30 maggio 1896 perché veda la luce il decreto ministeriale che attribuisce il titolo di libero docente in fisica sperimentale al giovane Dessau. Egli ha fatto dunque i conti con la lenta italica burocrazia, ma finalmente, dopo oltre due anni dall'inizio della procedura, ha in mano l'abilitazione che gli permette di candidarsi per avere incarichi di insegnamento. D'altra parte, ha ormai superato i trent'anni e si sta facendo urgente l'esigenza di trovare una sistemazione definitiva, sia sul fronte lavorativo che su quello affettivo/familiare. Dalla documentazione presente presso l'archivio dell'Università di Bologna, risulta che egli nel 1898 fa domanda di incarico presso alcuni istituti scolastici, come a esempio i regi licei di Udine e di Trapani. In effetti, nell'estate del 1898 riceve da quest'ultimo una proposta di assunzione per l'anno scolastico 1898/99. Dopo molte esitazioni, rinuncia all'incarico e rimane assistente a Bologna, ma viene ricompensato l'anno seguente con la nomina a direttore dell'Osservatorio Astronomico e Meteorologico universitario, con congruo aumento di stipendio (1900-1902). Inoltre, negli anni accademici 1902/1903 e 1903/1904 tiene, sotto la direzione del Righi, l'insegnamento di Fisica Elementare per gli studenti delle Scuole di Farmacia e Veterinaria, e quello della Meteorologia e Climatologia agli studenti della Scuola Superiore di Agraria, annessa all' Università di Bologna.

Va poi ricordato che proprio in quegli ultimi anni del secolo diciannovesimo nasce e si consolida rapidamente la Società Italiana di Fisica (SIF) e Augusti Righi è uno dei fondatori, attraverso la partecipazione al primo Consiglio Provvisorio insieme ai colleghi Battelli, Blaserna, Beltrami e Roiti [9]. Questo Consiglio, riunitosi a Roma, sotto la regia di Pietro Blaserna, il 5 agosto del 1897, prepara la bozza di statuto e convoca la prima assemblea per fine settembre, quando una cinquantina di partecipanti provenienti da tutta Italia, si incontrano, sempre a Roma, per approvare lo Statuto e costituire ufficialmente la Società Italiana di Fisica il cui scopo, come recita l'articolo 2 dello Statuto, è quello "di promuovere lo studio e il progresso della fisica". Oltre a esserne stato uno dei fondatori, Righi diverrà poco dopo vice-presidente e poi addirittura presidente della SIF ed è dunque del tutto verosimile pensare che Dessau, oltre a comparire tra i primi soci attivi, abbia avuto un ruolo importante nel coadiuvare il proprio maestro chiamato a ruoli di così grande responsabilità nella Società ([5]).

Intanto, nell'estate del 1900, durante una vacanza a Herrenalb nella foresta nera, Bernardo si fidanza con Emma Goitein, di quattordici anni più giovane, anch'essa discendente di una famiglia ebrea di tradizione rabbinica. Emma sarà sua moglie nel 1901 e rimarrà al suo fianco per oltre quarant'anni, fin quando cioè, gli orrori della guerra li costringeranno a una repentina e dolorosissima separazione, come sarà chiarito più avanti. Emma è figlia di Ida Löwenfeld (1848-1931) e del rabbino Gabor Goitein (1848-1883), di origini ungheresi ed esponente della neo-ortodossia. Dopo essere stato rabbino ad Aurich nel 1874, Gabor si trasferisce a Karlsruhe proprio nel 1877, quando nasce Emma [10]. Data la prematura morte del padre, Emma e i tre fratelli, Gertrud, Rahel ed Ernst, vengono cresciuti ed educati dalla mamma seguendo l'orientamento neo-ortodosso condiviso con suo marito e da giovani simpatizzano per il neonato movimento sionista [11]. Anche Bernardo, sulla scia della famiglia di sua moglie, sarà così coinvolto nel movimento sionista e prenderà parte attiva, nel ventennio successivo, alla vita dello stesso movimento in Italia. Emma si trasferisce dunque a Bologna nel 1901 e qui continua gli studi di arte e pittura, iniziati in Germania, presso l'Accademia di Belle Arti, divenendo una pittrice e ritrattista di pregio, specializzata nell'arte della xilografia. Emma, che è anche una valente musicista, riuscirà a conciliare questo impegno professionale e artistico di alto livello con il suo ruolo di madre ed educatrice fedele alla tradizione ebraica, dove spetta proprio alla donna di curare il legame tra la famiglia, come luogo degli affetti e delle tradizioni, e l'esperienza esterna di una società che, proprio in quegli anni, si sta rapidamente secolarizzando. Nel 1904 la famiglia di Bernardo ed Emma è allietata dalla nascita della primogenita Fanny. Per Bernardo è ormai improcrastinabile trovare una sistemazione lavorativa stabile ed economicamente soddisfacente, ma le occasioni di concorrere a una cattedra universitaria sono scarse e soggette a una dura competizione. Decide comunque di fare domanda per un posto di professore presso l'Ateneo Perugino, dal momento che c'è una cattedra vacante e Perugia non sembra una sede molto ambita. In effetti Dessau avrà un unico concorrente, vale a dire Guido Ercolini. La commissione composta da Blaserna, Roiti, Ruata, Bellucci e Battelli chiude i lavori a Roma il 23 novembre del 1904, verbalizzando la vittoria di Dessau con 46/50, un punto in più rispetto a Ercolini, come risulta dai verbali pubblicati nell'annuario 1904/1905 dell'Università di Perugia (riprodotto in [1], p.16). Nel medaglione che riassume i titoli di Dessau, si parla di 16 pubblicazioni presentate, ma, tra lavori di rassegna, traduzioni e commenti, la commissione non può fare a meno di notare che ci sono solo *"tre lavori sperimentali, di cui il più importante è quello in cui studia gli strati metallici prodotti dal disgregamento di catodi, per effetto delle scariche elettriche"*. Insomma, il lavoro migliore, che gli permette di salire in cattedra, resta quello relativo alla tesi di laurea pubblicata sugli Annalen der Physik und Chemie nel 1886, ancor prima di giungere in Italia. Torneremo più avanti ad approfondire la questione della scarsa produzione bibliografica del periodo bolognese.

---

([5]) Purtroppo, non è stato possibile rintracciare la lista con i nomi dei primi soci della SIF, tuttavia è ragionevole supporre che Dessau, ormai libero docente per il sistema italiano, abbia aderito alla SIF sin dalla sua Fondazione. A partire dal 1905, egli aderirà anche alla Società Tedesca di Fisica.

**2.3 Gli incontri con il giovane Marconi e la misteriosa deposizione notarile del 1938**

In relazione al periodo bolognese dei primi anni '90 e al lavoro sulle onde elettromagnetiche, è interessante sottolineare il ruolo di Dessau come testimone diretto degli incontri tra Righi e Marconi, del quale si ricordano proprio in questo anno 2024 i centocinquanta anni dalla nascita. Del resto, la questione dei rapporti tra questi due giganti della storia della fisica italiana, in particolare dell'influenza del lavoro di ricerca di Righi sullo sviluppo delle invenzioni di Marconi, è stata oggetto di interesse fin dai tempi in cui si svolsero i fatti e sono apparsi di recente interessanti approfondimenti ([4] pp. 215-228 e [12]). Dessau parla esplicitamente delle visite di Marconi al laboratorio di Fisica presso l'Università di Bologna nel discorso che, come sarà ripreso più avanti, tiene nel 1907 al convegno organizzato dalla SIF a Roma per celebrare i primi venticinque anni di ordinariato di Righi e così si esprime: [13] *"Sappiamo che le sue (di Righi, ndr) ricerche sulle onde elettriche ebbero pure un risultato di somma importanza pratica, la possibilità del quale si era rivelata al geniale intuito del giovane Marconi, allorché, me presente, ebbe occasione di conoscere nel* *laboratorio di Righi gli apparati da questi ideati e gli scopi a cui dovevano servire. Infatti, il primo apparecchio produttore di onde adoperato dal Marconi per la sua telegrafia senza filo non è altro che l'oscillatore del Righi, e se è giusto che il mondo tributi al Marconi l'onore dovuto all'audace sua iniziativa ed alla meravigliosa sua perspicacia e abilità scientifica, noi altri fisici dobbiamo vieppiù insistere sul sommo merito dell'indagine scientifica".*
Dessau sembra quindi convinto, basandosi sulla sua diretta esperienza, di quanto Marconi fosse stato influenzato e ispirato dagli apparati visti nel laboratorio di Righi. Questa opinione, del resto, era già stata implicitamente espressa quattro anni prima da Dessau e Righi nell'introduzione, scritta a quattro mani, del volume La Telegrafia senza Filo [14] *"Il lettore terrà conto delle difficoltà che dovevano superarsi, fra le quali quella che dipende dal fatto che non di rado qualche inventore di professione ottiene brevetti per metodi o strumenti già noti nella scienza, qualche volta dopo avervi introdotto tutt'al più delle piccole modificazioni."*. L'allusione al Marconi sembra qui evidente, anche perché nel capitolo dedicato alla generazione delle onde elettromagnetiche, firmato da Dessau, si conferma in modo esplicito l'idea che Marconi avesse inizialmente utilizzato un generatore di onde identico a quello di Righi: *"Infatti, se all'apparecchio produttore delle onde descritto da Marconi nel suo primo brevetto…. confrontiamo quello a tre scintille del Righi… riscontriamo tra esse un'identità perfetta"* (Ref. [14] p. 287). D'altra parte, poche righe più avanti, Dessau riconosce onestamente e pienamente a Marconi *"il merito indiscutibile di aver preso una audace iniziativa, laddove da altri non erano state fatte che delle timide proposte, e di aver trasportato nel campo pratico ciò che altri avevano soltanto intravveduto o realizzato in scala minore. Ma l'ingegno e le facoltà inventive sue si rivelarono pienamente più tardi; per l'abilità colla quale vinse le numerose difficoltà, e per le tante modificazioni ed aggiunte di dettaglio in gran parte essenziali per il successo pratico, che furono da lui introdotte e riunite in quell'insieme, che a ragione può chiamarsi il sistema Marconi"* ([6]).
Da queste testimonianze dei primi del novecento possiamo dedurre quindi come Dessau avesse personalmente assistito a diverse visite del giovane Marconi presso i laboratori dell'Istituto di Fisica bolognese. Addirittura, sapendo come siano fitte di appuntamenti le agende dei professori del calibro di Augusto Righi, sovraccarichi di impegni e responsabilità, si potrebbe supporre che il Righi potesse aver delegato il suo giovane assistente a illustrare all'ancora semi-sconosciuto giovane Marconi i primi apparati ed esperimenti sulle onde elettromagnetiche. Ma il ruolo di Dessau in questi incontri viene ridimensionato dalla testimonianza verbalizzata dal notaio Alberto Tei di Perugia nel settembre 1938 (Fig. 5). Il testo di questa deposizione, che era stato già parzialmente incluso in Ref. [15], è riportato qui di seguito.

---

([6]) E' anche interessante notare che dalle lettere tra Marconi e Righi conservate presso la biblioteca Bodleian di Oxford, risulta che Marconi, ai primi di maggio del 1903, scrive dall'Inghilterra una lettera raccomandata a Righi per lamentarsi di quanto contenuto nell'appendice del volume "La Telegrafia senza Filo", firmata da Dessau e dedicata a "Le recenti esperienze a distanza grandissima". Righi, dopo aver chiarito che la responsabilità di quelle pagine è "del mio collaboratore", cioè di Dessau, cerca comunque di giustificarlo, spiegando che egli ha soltanto riportato, con tono comunque dubitativo, delle affermazioni fatte da un certo Maskelyne su fatti che poi, solo in un secondo tempo, si sarebbero dimostrati falsi. Assicura quindi Marconi che nella nuova edizione che è in preparazione, Dessau apporterà le necessarie rettifiche e correzioni.

*"Repertorio 17439 - Deposizione testimoniale.*
*L'anno millenovecentotrentotto, il giorno venerdì sedici di settembre, in Perugia… davanti a me Dottore Alberto Tei, notaio, registrato a Perugia, iscritto al collegio notarile del distretto di detta città, senza l'esigenza dei testimoni per avervi il comparente, col mio consenso, espressamente rinunziato, è presente*

*L'illustrissimo Signor Professore Bernardo Dessau, fu Samuele, nato a Offenbach (Germania), domiciliato a Perugia, già docente di fisica presso la R. Università di Perugia di piena capacità giuridica…., all'oggetto di stipulare il presente atto, …, a richiesta del Sig. Ingegnere Aldo Righi, al solo scopo di voler fare risultare alcune circostanze e fatti risalenti al periodo del suo assistentato presso il Prof. Augusto Righi di Bologna, fatti e circostanze che egli è disposto a confermare, anche sotto il vincolo giuramento, avanti a qualsiasi autorità e cioè:*

*Negli anni 1894 o 1895 il prof Augusto Righi si occupava a realizzare con le onde elettriche degli effetti e fenomeni ottici e di porre così su piena base sperimentale quell'intrinseca identità tra i due gruppi di fenomeni, che Enrico Hertz non aveva ancora potuto dimostrare completamente. Le onde corte all'uopo necessarie, il Righi le ottenne con un apparecchio di sua creazione, detto oscillatore, che veniva collocato lungo la linea focale di uno specchio metallico di forma di cilindro parabolico; ed uno specchio analogo raccoglieva poi le onde dopo un risonatore lineare collocato sulla sua linea focale.*

*Appartengono a quell'epoca le visite del giovane Marconi al Prof. Righi. Io stesso conobbi il Marconi trovandomi in uno degli ambienti attigui all'aula di fisica: ambienti nei quali io di solito lavoravo e dove egli aspettava di essere ricevuto dal Professore. Ma le visite di Marconi al Righi avvenivano nel laboratorio di quest'ultimo, una sala alla quale si accedeva, oltre che dal lato dell'aula, anche più direttamente salendo la scala dell'ingresso di Via Zamboni 31. Alle visite del Marconi in quella sala, io non sono stato mai presente, essendo occupato più che altro nella preparazione delle esperienze per le lezioni, nelle esercitazioni degli studenti e nelle ricerche dei laureandi. Ma il giovane Aldo, figlio del Righi, che in quegli anni io vedevo spesso venire a passare delle ore nel laboratorio del padre, ricorderà certamente di aver visto il Marconi. Dalla bocca del Prof. Righi io sapevo che Marconi veniva a parlargli di invenzioni e progetti suoi, e poiché nella sala dove Righi lo riceveva, stava bene in vista tutta l'apparecchiatura per lo studio delle oscillazioni e onde elettriche, l'attenzione del giovane Marconi non può non essere stata colpita da quegli apparecchi che egli deve averne chiesto schiarimenti al Righi.*

*Per bocca del Righi ho saputo anche di visite del Marconi alla villa di Sabbiuno di Montagna, dove egli in quegli anni passava l'estate con la famiglia e che sovrasta Villa del Grifo, giù nella valle del Reno dove abitava Marconi. Anche queste visite, così mi disse il Righi, erano dedicate alle conversazioni che il giovane desiderava d'avere con lui intorno ai suoi progetti.*

*Ricordo infine che una volta venne in laboratorio per incontrarsi con Righi, come non di rado faceva, il Prof Tizzoni, anche lui dell'Università di Bologna, Deputato al Parlamento, egli era tornato da Roma dove aveva assistito alle dimostrazioni che il Marconi, reduce dai primi suoi trionfi in Inghilterra, aveva dato dell'opera sua davanti a personaggi del mondo ufficiale e politico. E il Tizzoni, introdotto da me nella sala di lavoro di Righi per aspettare la venuta di quest'ultimo, davanti agli apparecchi del Righi proruppe in un'esclamazione di*

*grande sorpresa di trovarsi già in quell'ambiente gli apparecchi che egli riteneva quelli di Marconi, invece erano quelli di Righi.*

*Ecco quello che sono in grado di contribuire a definire della posizione del Prof. Righi sulla nascita della telegrafia senza fili."*

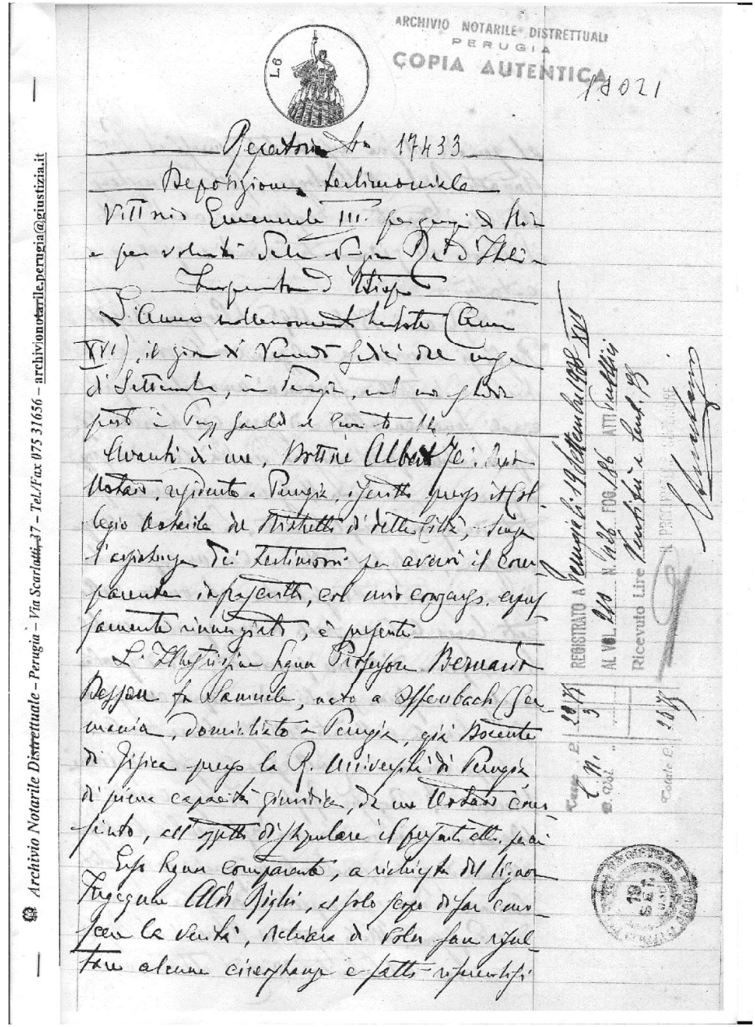

**Fig. 5** Prima pagina della deposizione notarile del 16 settembre 1938, resa da Dessau a proposito dei rapporti tra Righi e il giovane Marconi negli anni novanta dell'ottocento. Archivio Notarile Distrettuale di Perugia.

Risulta difficile comprendere il motivo per cui Dessau, ormai settantacinquenne e a distanza di oltre quarant'anni dai fatti narrati, senta il bisogno di depositare presso un notaio questa testimonianza formale, che in un certo senso contraddice le sue affermazioni rese pochi anni dopo i fatti in questione, quando aveva dichiarato di essere stato personalmente presente agli incontri di Righi col giovane Marconi. Si potrebbe pensare che, essendo morto il Marconi nel 1937, potessero essere emerse controversie o rivendicazioni di tipo giuridico per questioni di eredità o di brevetti che siano all'origine di questa testimonianza. Ma, per quanto ci è stato possibile indagare, nulla di tutto ciò è emerso, anche perché le tecnologie utilizzate all'epoca dei fatti erano ormai del tutto superate negli anni trenta. Piuttosto, ci sembra di poter ricondurre questo episodio alla volontà dell'Ing. Aldo Righi di tenere viva la memoria del padre e fare in un certo senso giustizia ai suoi meriti, dal momento che i successi di Marconi (premio Nobel incluso) hanno oggettivamente messo in secondo piano, almeno per il grande pubblico, il ruolo fondamentale di Righi nella realizzazione dei primi apparecchi per le onde elettromagnetiche. Questa interpretazione è anche rinforzata dal fatto che, più o meno nello stesso periodo, l'ing. Aldo Righi sollecita anche un'altra deposizione sui contatti tra il giovane Marconi e Augusto Righi nel periodo 1894/1895. Si tratta della testimonianza del Conte Giuseppe Manzoni-Ansidei, resa il 3 Luglio 1939 davanti al notaio Antonio Magli del Collegio Distrettuale Notarile di Bologna, dove si ricordano le esperienze di trasmissione delle onde elettromagnetiche mostrate da Righi agli studenti nel 1895. Il testimone conclude quindi in questo modo: *"Certamente Guglielmo Marconi che mi dissero che assistesse a quelle lezioni, come era notorio che frequentasse il gabinetto di fisica dell'Università ed avesse frequenti incontri con il Righi, avrà fatto tesoro di quelle ricerche del grande Maestro che in quelle lezioni parve divinare il telegrafo senza fili e anche la radio* [Ref. 4, p. 219]. Insomma, anche in questo caso ci sembra evidente che lo scopo della raccolta di queste testimonianze fosse quello di rivendicare, da parte di discendenti della famiglia Righi, come all'origine delle idee e delle invenzioni del Marconi, addirittura fino alla radio che sarebbe apparsa quasi trent'anni dopo, ci fossero le idee e gli apparati messi a punto da Augusto Righi.

Del resto, anche sulla questione dell'assegnazione del premio Nobel a Marconi nel 1909 ([7]), gli stessi Righi e Dessau si dimostrano da subito molto perplessi, come risulta dalla loro corrispondenza [7]. Nell'apprendere la notizia, Dessau, che si trova in quelle settimane di dicembre del 1909 a Mannheim, scrive a Righi di essere rimasto sfavorevolmente colpito dal risultato che ha avuto il conferimento del Nobel e lo assicura che, con tutte le precauzioni, ne parlerà a Berlino con Rubens e Warburg e, se potrà, con Röntgen a Monaco. Subito dopo le vacanze di Natale, scrive di nuovo che ha effettivamente parlato con Rubens e altri professori tedeschi, che hanno espresso la loro stima nei riguardi dei meriti scientifici di Righi, augurandosi che esercitino la loro influenza in occasione della prossima assegnazione del Nobel. In realtà, come già accennato sopra, malgrado le quindici candidature, Righi non conseguirà mai l'ambito premio.

**2.4 Un bilancio del periodo bolognese**

Provando a soppesare, con il dovuto distacco, la grossa mole di lavoro svolto da Dessau nel periodo bolognese, non si può fare a meno di notare la sproporzione rispetto al magro bottino, soprattutto in termini di visibilità scientifica e di pubblicazioni, che egli è riuscito ad accumulare. Come sottolineato in precedenza, persino la commissione del concorso perugino non poté fare a meno di mettere a verbale che tra i titoli presentati dal candidato, ormai quarantunenne, c'erano sono solo tre lavori sperimentali, il migliore dei quali era la tesi di laurea fatta a Strasburgo. D'altra parte, guardando alla produzione scientifica del suo mentore nello stesso periodo, colpisce che durante l'ultimo decennio dell'800, cioè dal 1890 al 1899, Augusto Righi pubblica una settantina di lavori, tra articoli e comunicazioni a conferenze, cioè una media di sette lavori all'anno, che per l'epoca doveva essere davvero un record bibliometrico. Ebbene, nello stesso arco temporale, per quanto ci è stato possibile ricostruire (una dettagliata bibliografia è inclusa in Ref. [1] pp. 42-47), Bernardo Dessau risulta autore della traduzione in tedesco del testo del Righi sull'Ottica delle Oscillazioni Elettriche [5], due traduzioni e un articolo di rassegna, oltre a un paio di lavori sperimentali sugli atti dell'Accademia dei Lincei [16, 17]. E' chiaro che non possiamo in alcun modo proiettare sul passato i criteri bibliometrici che adotteremmo oggi per giudicare come un docente universitario faccia maturare il suo allievo verso la carriera accademica. Colpisce però quanto poco abbia pubblicato in quegli anni il Dessau, malgrado fosse arrivato in Italia da neolaureato, con la tesi pubblicata sugli Annalen der Physik un Chimie. Ad esempio, sembra lecito domandarci perché Dessau non sia stato coinvolto come co-autore del trattato di Righi sull'ottica delle oscillazioni elettriche del 1898, o almeno citato in dei ringraziamenti all'interno di tale opera, visto che essa è appunto il frutto degli esperimenti e degli studi svolti dal Righi proprio a partire dall'arrivo a Bologna nel 1890 insieme al suo assistente. Questo deficit di visibilità e di produzione scientifica, sarà pagato da Dessau nella sua successiva carriera accademica, dal momento che, a parte il successo occasionale e inatteso per la sede poco ambita di Perugia, non riuscirà più a vincere alcun concorso per sedi più prestigiose e attrezzate, a cui pure si dimostrerà molto interessato. Tra le cause di questo che oggi classificheremmo come un clamoroso flop bibliometrico, va probabilmente anche considerato il carattere mite e per nulla ambizioso del Dessau, unito alla titubanza nel decidere se e dove proseguire la carriera accademica. Del resto, per rendersi conto di quanto la sua scala di priorità fosse un po' diversa da quella dei suoi colleghi dell'epoca, basti ricordare che egli considerava importantissimo l'accordo stretto con Righi di avere garantita la possibilità di osservare il riposo sabbatico. Nella prospettiva di Bernardo, questa condizione è risultata fondamentale per prolungare forse oltre il dovuto la pur precaria sistemazione bolognese, dal momento che, trasferendosi altrove, questa sorta di privilegio riguardo all'osservanza dello Shabbat avrebbe potuto essere messa in discussione.

---

([7]) Marconi, che non si era neanche mai iscritto all'Università, ottenne il Nobel nel dicembre 1909 per i suoi contributi allo sviluppo della telegrafia senza filo, sull'onda della felice vicenda del salvataggio di tutti i circa millesettecento passeggeri del transatlantico Republic, che era stato speronato al largo delle coste nordamericane nel gennaio del 1909, grazie al messaggio lanciato via radio dal marconista a bordo della nave.

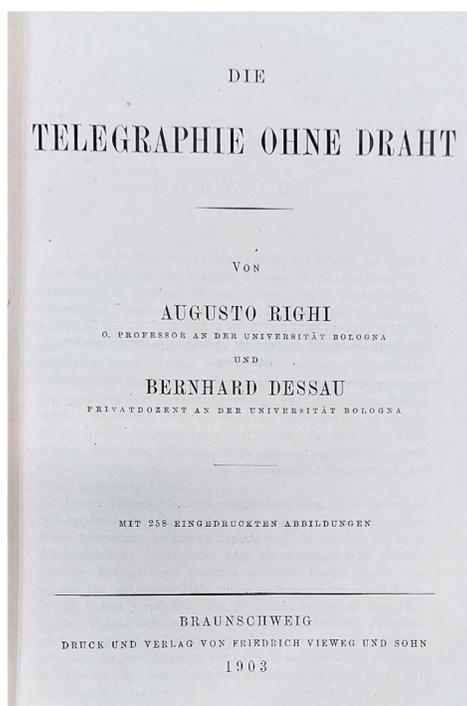

**Fig. 6** Copertina dell'edizione tedesca de "La Telegrafia senza Filo" edito in Germania (collezione del Dipartimento di Fisica e Astronomia "Augusto Righi" di Bologna.)

### 3. Il trentennio accademico perugino (1905-1935)

#### 3.1 I primi quindici anni a Perugia, tra impegno sionista e voglia di trasferirsi altrove.

Come abbiamo già anticipato in precedenza, il 1904 segna una svolta nella vita di Dessau. La famiglia è allietata dalla nascita della primogenita Fanny e, soprattutto, egli vince il concorso per la cattedra di Fisica Sperimentale presso l'Università di Perugia. Il suo trasferimento dovrebbe avvenire all'inizio dell'anno accademico 1904/1905, ma, su richiesta scritta di Righi (riprodotta in [1] a p. 59), i rettori di Bologna e Perugia si accordano sul posticipo del trasferimento, per dar modo all'Università di Bologna di organizzarsi per supplire all'assenza di Dessau. Il trasloco avviene quindi nel gennaio 1905 e, solo a inizio febbraio, egli avvia le lezioni presso il suo nuovo ateneo. Per lui ed Emma non è facile adattarsi al cambiamento: da una città viva e ben collegata come Bologna a una piccola provincia isolata nel cuore verde d'Italia. Come annoterà la figlia Fanny molti anni dopo, *"Allora Perugia era veramente la «Città del silenzio», adagiata sui suoi colli in mezzo alla vasta chiostra di monti ed agli sconfinati orizzonti…. Ricca di tesori d'arte e di tradizioni antichissime, Perugia stava in disparte, lontana dalle vie di comunicazione ed anche dal progresso del mondo"* [18]. Il disagio e la difficoltà di adattamento della giovane coppia all'ambiente perugino durerà quasi un ventennio, ma alla fine essi faranno pace con questa città che sentiranno finalmente come loro patria e che riserverà loro amicizia e perfino salvezza nei tempi scuri a venire.

In quegli anni di inizio secolo, Bernardo diventa parte attiva nel nascente movimento sionista che si sviluppa in Europa in seguito ai disordini antiebraici (progrom) in Ucraina ed in Russia meridionale dei primi anni ottanta dell'ottocento, che hanno segnato un deciso salto di qualità nelle conseguenze dell'antisemitismo serpeggiante da sempre sia nell'est che nell'ovest del vecchio continente. Grazie all'influsso della famiglia di Emma, di origini ungheresi, la famiglia Dessau viene quindi da subito coinvolta nelle attività del movimento sionista, fondato dall'ungherese Theodor Herzl (1860-1904) durante l'ultimo quinquennio dell'ottocento. Nei primi anni del nuovo secolo, Bernardo partecipa a diversi congressi dei Sionisti Italiani, mentre nel 1903 e 1905 si reca addirittura a Basilea come delegato italiano al Sesto ed al Settimo Congresso Sionista Mondiale [19, 20]. Nel corso del IV Congresso sionistico italiano (1904), presenta una relazione sul tema *"Atteggiamento dei sionisti italiani di fronte alle condizioni degli ebrei in Oriente"* (l'intervento fu pubblicato poi su "L'Idea Sionnista" del marzo-aprile-maggio 1904). In occasione dell'VIII Congresso sionistico italiano (1907) scrive un saggio dal titolo: *"Il primo decennio del Movimento sionnista in Italia"*, uscito in quell'anno nel volumetto *"L'8° congresso sionnista. I risultati dell'8 Congresso"* a cura di Felice Ravenna (si noti che in quegli anni si usava spesso la doppia "n"). Nell'ambito di questo impegno sionista, la famiglia Dessau diventa un imprescindibile punto di riferimento per la piccola comunità ebraica perugina [21] ed essi si rendono disponibili, in quegli anni, ad accogliere in casa loro alcuni rifugiati ebrei scappati dall'Europa orientale. Essi sono anche tra gli animatori della rivista "L'Idea sionista", nata nel 1901 per iniziativa di Carlo Conigliani di Modena e stampata regolarmente fino al febbraio 1911. Morto il direttore Conigliani dopo il primo anno di pubblicazione, la direzione viene affidata a Benvenuto Donati per un biennio e poi al modenese prof. Carlo Levi, che rimane fino al termine delle pubblicazioni e decide anche di mutare il nome in "Idea Sionnista", aggiungendo la doppia "n". Se nei primi anni di vita della rivista Dessau contribuisce con alcuni articoli, nel 1906 ha luogo un tentativo di riorganizzazione ed egli viene addirittura proposto da Angelo Sullam come possibile nuovo direttore con queste motivazioni: *"Quanto al Direttore, esso non può e non deve essere che Dessau, l'unico che possa far risorgere l'Idea Sionnista. Dessau è in buoni rapporti con il Corriere Israelitico, è sionista nazionalista moderato, è molto calmo, molto abile, molto persuasivo"* ([20], p.66). Alla fine, Bernardo rimane semplice redattore, mentre Emma cura le illustrazioni e la copertina a partire dal 1906.

Non è questa la sede per analizzare nei dettagli la posizione di Dessau all'interno del movimento sionista, dal quale poi si allontanerà negli anni venti. Basti qui ricordare che lui sostiene con forza, in quegli anni, la tesi che il movimento dovesse avere carattere non tanto politico, economico o filantropico, quanto piuttosto identitario e nazionale, nel senso di riallacciare i legami con le generazioni e le tradizioni del passato:

*"Scientificamente si potrà forse dimostrare (non entro nell'argomento) l'origine economica del movimento; ma fatto sta che la grande massa degli ebrei non lo vuole considerato che come un risorgimento nazionale, dal quale naturalmente non va disgiunto il risorgimento economico. [...] Donde per noi l'obbligo di comprendere e di assistere il movimento nel suo vero carattere. [...] A mio parere, chi pretende o chi crede di essere sionista unicamente per motivi filantropici, s'illude semplicemente. Scrutando bene i propri sentimenti, troverà sempre di essere legato colla grande massa del popolo ebreo da qualche cosa di più di un semplice sentimento di compassione. [...] So benissimo che vi sono degli ebrei, in Italia come altrove, pei quali questi legami realmente non esistono più, perché purtroppo furono rotti già da molto tempo. Ma se la generazione attuale non ne ha colpa, una colpa c'è stata lo stesso, e fu di coloro i quali avevano ancora conosciuto le nostre tradizioni e non sentirono per esse, per un passato che è quello dei loro avi, un affetto sufficiente per tener care quelle tradizioni e per infonderne l'amore anche ai figli. Ed allora, a mio debole parere, sarebbe dovere dei figli di riparare l'errore degli avi, e di ricordarsi del loro passato, anziché cercare di dimenticarlo completamente."* [Ref. [20], p. 62].

È anche interessante notare, tuttavia, come Dessau chiarisca esplicitamente che la fedeltà alle tradizioni religiose ed etniche ebraiche debba andare di pari passo con la sincera fedeltà alla patria italiana:

*A noi altri ebrei dell'occidente, educati assieme ai figli delle nazioni tra le quali viviamo, imbevuti di idee moderne, riesce difficile comprendere come vi possano essere delle tradizioni comuni, se non sono quelle religiose, in un popolo che da 1000 anni ha perduto il proprio paese e l'indipendenza, [...]. Eppure, tra quelle popolazioni, la continuità della tradizione nazionale, anche nell'esilio, non venne mai interrotta. [...] [...] l'ebreo italiano deve e può essere sinceramente fedele alla patria italiana, pur sapendo di far parte, assieme*

*all'ebreo russo, di una medesima grande unità etnica, e di avere comune con questo la tradizione semitica."* (intervento al quarto congresso sionistico italiano [20], p. 252).

Dal punto di vista scientifico, poco prima del trasferimento a Perugia, vede finalmente la luce un'opera a cui Dessau sta lavorando da parecchi mesi insieme a Righi: la "Telegrafia senza filo" [14] che viene poi anche pubblicata in tedesco con il titolo "Telegraphie ohne Draht" (copertina riprodotta in Fig. 6) [22]. Successivamente uscirà una seconda edizione riveduta e ampliata sia in italiano [23] che in tedesco e [24]. Si tratta di un'opera corposa e approfondita, dove, a parte l'introduzione scritta a quattro mani, ciascuno degli autori firma alcuni capitoli, mentre Emma si occupa di illustrare la copertina. Non può sfuggire che i capitoli di carattere introduttivo e fondamentale sono appannaggio del Righi mentre quelli più tecnici, che entrano anche nei dettagli delle tecnologie usate da Marconi per le sue trasmissioni di onde radio, siano firmati proprio da Dessau. Negli anni bolognesi, del resto, egli si era occupato direttamente di questa tipologia di esperimenti, come attestano anche le scritte in tedesco che accompagnano diversi coesori (coherer) ancora visibili presso il Dipartimento di Fisica e Astronomia di Bologna e di cui una immagine è riportata in [25]. L'opera ha talmente successo che, come accennato sopra, nel 1907 uscirà la seconda edizione. Nello stesso anno nasce Gabor, il secondogenito della famiglia Dessau.

Ma il 1907 è anche l'anno in cui a Bernardo vengono chiesti interventi di rilievo per la comunità scientifica sia perugina che nazionale. Anzitutto, egli è incaricato di tenere il discorso inaugurale dell'anno accademico presso Università di Perugia e lui sceglie di intitolarlo "L'evoluzione della Materia", poi stampato nell'annuario dell'Università di Perugia [26]. Leggendo questo discorso si può apprezzare non solo lo stile elegante e sobrio dell'eloquio, ma anche come egli sia in grado di rappresentare, a un pubblico di non specialisti, la fase davvero critica e gravida di novità che la fisica attraversa agli inizi del secolo, soprattutto in relazione alla evoluzione dalla fisica classica a quella quantistica e alla comprensione della struttura atomica della materia. Addirittura, come ha notato anche il chimico perugino prof. Reichenbach ([1], p .15) *"dà descrizioni della struttura dell'atomo, ..., del decadimento radioattivo e delle grandi energie connesse ad esso"*.

Nello stesso anno 1907, è affidato a Dessau, dal presidente Vito Volterra, un compito di primissimo piano sul palcoscenico nazionale: tenere la relazione commemorativa al convegno organizzato dalla SIF a Roma il 13 marzo per celebrare il venticinquennale della cattedra di Augusto Righi [13]. Successivamente, quando il 12 aprile si tengono a Bologna le solenni celebrazioni per il giubileo di Augusto Righi e la contemporanea inaugurazione della nuova maestosa sede dell'Istituto di Fisica, Volterra cita Dessau in questo modo durante il discorso rivolto a Righi: *"Pregato da noi, il Vostro allievo diletto, il collaboratore Vostro infaticabile di tanti anni, Bernardo Dessau, compose uno scritto sulla Vostra opera scientifica, che commosso ci lesse nella seduta del 13 marzo. Nella loro sobria ed affettuosa semplicità, le pagine del Dessau molte cose nuove ci han rivelato, molte ci han ricordato che non debbono restare nell'oblio; in quelle pagine l'intera opera Vostra, concatenata nelle sue varie parti, grandiosa nel suo insieme, feconda nelle sue conseguenze, spicca come splendido monumento del Vostro ingegno e del vostro sapere"* [27].

In effetti il discorso di Dessau del 13 marzo viene preparato scrupolosamente, tanto che nelle settimane precedenti egli scrive a più riprese allo stesso Righi [7] per chiedergli notizie di prima mano sulla sua iniziale vocazione di fisico e su altri passaggi poco noti della sua carriera. Dessau narra quindi con dovizia di particolari dell'orientamento del Righi verso i fenomeni elettrici fin da giovanissimo, degli studi di matematica e della laurea in ingegneria.

Ricorda poi gli studi giovanili del Righi volti alla costruzione del "telefono che si ascolta a distanza", allo studio dell'isteresi magnetica e alla macchina elettrostatica (precursore di quella di Van der Graaf). Racconta quindi dell'incarico di insegnante all'istituto tecnico di Bologna e del trasferimento nel 1880 come docente universitario a Palermo, dove approfondisce gli studi sull'effetto Hall anomalo nel bismuto e sull'effetto Kerr magnetoottico. Segue il trasferimento a Padova nell'anno scolastico 1886/87 dove si dedica all'ottica dei materiali anisotropi, ma soprattutto all'effetto fotoelettrico di cui di fatto è lo scopritore in contemporanea con W. Hallwachs, sebbene solo quest'ultimo ne abbia avuta riconosciuta la paternità. Nell'autunno del 1889 ha la cattedra a Bologna, dove continua lo studio dei fenomeni elettrici nei tubi a scarica (ecco perché gli serve chi soffi il vetro), inclusa la novità del momento riguardante i raggi x, per poi dedicarsi intensamente per diversi anni allo studio sistematico delle oscillazioni elettriche, perfezionando e approfondendo il lavoro lasciato in sospeso da Hertz.
Sempre a partire dall'anno 1907, Dessau si iscrive anche alla Società Italiana per il Progresso delle Scienze, che era nata a Pisa nel 1839, riunendo ben 421 scienziati provenienti dai vari regni in cui era ancora divisa la penisola. La società rimane sostanzialmente dormiente dal 1875 fino al 1906, quando viene definitivamente ricostituita adottando un nuovo statuto, nato dal prepotente bisogno di promuovere nel Paese il progresso, la coordinazione e la diffusione delle scienze. La ricostituita Società, nel primo decennio del secolo XX tiene la I, la II e la III riunione rispettivamente a Parma (1907), Firenze (1908) e Padova (1909) sotto la presidenza del celebre fisico-matematico Vito Volterra che sarà anche il fondatore e primo presidente del CNR. Proprio al congresso di Padova nel settembre del 1909 ([1], p. 21), Dessau presenta la relazione intitolata "Masse e dimensioni degli elementi costitutivi della materia" dove analizza l'interpretazione microscopica della termodinamica e spezza più di una lancia a favore di quella che chiameremmo oggi teoria cinetica dei gas, concludendo che:
*"un secolo dopo Dalton, il concetto atomistico della materia, da geniale ipotesi, è passato nel campo delle convinzioni scientifiche meglio assodate"* [28].

A dispetto dell'aumentata visibilità del Dessau sul palcoscenico nazionale, l'attività dedicata alla ricerca sperimentale langue, non potendo disporre, presso le scuole di medicina e di farmacia di Perugia, di un vero e proprio laboratorio per svolgere ricerche avanzate. Non resta quindi che dedicarsi a pubblicare traduzioni o articoli di rassegna. Nel 1905 e poi nel 1908 Dessau traduce in tedesco le due edizioni del libro di Augusto Righi sulla "Moderna Teoria dei Fenomeni Fisici" [29], mentre nel 1910 egli dà alle stampe un volume di rassegna, in lingua tedesca, sulle proprietà chimico-fisiche delle leghe metalliche "Die physikalisch - chemischen Eigenschaften der Legierungen" [30]. Dal momento poi che la principale attività perugina è quella della didattica della fisica sperimentale agli studenti di medicina e farmacia, Bernardo si pone il problema di preparare un buon libro di testo. Nel 1912 esce dunque per i tipi della Società Editrice Libraria di Milano il primo volume del suo "Manuale di Fisica" dedicato alla Meccanica. Nel giro del quinquennio successivo vedranno la luce il secondo volume, dedicato ad Acustica, Termologia e Ottica e il terzo volume dedicato all'Elettrologia. Una seconda edizione dei tre volumi apparirà nel 1928 e addirittura la terza edizione del terzo volume, dove si recepiscono alcune delle principali novità della nascente fisica quantistica, apparirà nel 1935. Sarà uno dei manuali di fisica più diffusi in Italia ed avrà successo anche all'estero, grazie all'edizione in lingua tedesca, segnando un impegno di ottimizzazione e di aggiornamento continuo, durato più di vent'anni. Particolarmente apprezzabili, ancora oggi, mi sembrano soprattutto le parti dedicate alla spiegazione dettagliata del funzionamento di tutta una serie di apparati e macchine, come ad esempio i tubi sonori o i diapason, gli spettrometri ottici, le macchine elettrostatiche, per non parlare della corposa e dettagliata sezione di elettrotecnica.

Si deve sottolineare che in questi primi anni di permanenza a Perugia, Bernardo non rinuncia al sogno di trasferirsi presso sedi più attraenti o attrezzate per il lavoro sperimentale, contando magari sull'appoggio di Righi, come si rintraccia facilmente nella loro corrispondenza epistolare [7]. Ad esempio, nel maggio 1908 Dessau scrive a Righi dicendogli di aver saputo che Corbino si trasferirà da Messina a Roma e che quindi ci sarà presto un concorso per la cattedra vacante di Messina. Chiede dunque a Righi di interessarsi della cosa e possibilmente di manifestare la propria disponibilità a far parte della commissione che si dovrà eleggere a giorni, in modo da avere la garanzia di partecipare a un concorso nel quale non passerebbero eventuali favoritismi e nel quale saprebbe di essere sostenuto in modo lecito dal suo mentore. Nell'ottobre del 1912 scrive ancora a Righi riferendogli che parteciperà al concorso per la cattedra di fisica sperimentale al Politecnico di Torino e pregandolo, qualora conosca qualche membro della commissione, di intervenire in suo favore. Il mese successivo, scrive di nuovo dichiarando di aver saputo che proprio Righi è stato nominato a far parte della commissione per il concorso di Torino, augurandosi che sia disposto a garantire anche in questo caso che la sua candidatura sia presa in considerazione. Nessuno di questi tentativi sarà coronato da successo, anche perché, come accennato in precedenza, i titoli scientifici di Dessau sono piuttosto limitati, a causa essenzialmente della aridità del periodo bolognese. L'ultimo tentativo di lasciare Perugia Dessau lo compie nel 1921, quando, ormai alla soglia dei sessant'anni, pensa addirittura di trasferirsi in Palestina (all'epoca sotto il Mandato Britannico), dove si sta formando il focolare nazionale ebraico. Dessau scrive nientemeno che ad Albert Einstein ([1], p.25), ormai salito alla ribalta della fama internazionale, chiedendogli un consiglio sull'eventualità di ottenere una posizione presso il Technion di Haifa. La risposta di Einstein [31] arriva da Berlino nel marzo successivo (con ben sei mesi di ritardo, di cui Einstein si scusa a motivo del suo viaggio in America) ed è alquanto scoraggiante, perché sostiene che *"andare ad Haifa non può essere consigliato a un fisico affermato che non sia del tutto posseduto dall'idealismo"*, dal momento che si tratta di una semplice università didattica rivolta agli studi tecnici e di ingegneria. Piuttosto, Einstein suggerisce di puntare alla nuova Università di Gerusalemme, in via di costruzione, che potrebbe ospitare anche un istituto di fisica se avrà successo la campagna di raccolta fondi per favorire la quale Einstein è stato in America proprio di recente. In effetti i tempi si dimostrano non ancora maturi e il reclutamento di eminenti fisici europei presso l'Università di Gerusalemme si materializzerà soltanto a partire dagli anni trenta. Anche il trasferimento in Terrasanta rimarrà dunque un sogno nel cassetto.

Dopo un decennio di apprezzato servizio come professore straordinario presso l'ateneo perugino, nel 1913 Bernardo Dessau ottiene la promozione a ordinario. Nemmeno il tempo di gioire per questa stabilizzazione, che tornano problemi di salute per cui, nel 1914, lo scoppio del primo conflitto mondiale lo trova in Germania dove si è recato per sottoporsi a un delicato intervento chirurgico. Alla notizia dell'inizio della guerra, torna precipitosamente in Italia e deve sottoporsi a una nuova operazione, visto che la precedente non ha avuto l'esito sperato. A quel periodo si riferisce il ritratto con la figlia Fanny, realizzato a carboncino dalla moglie Emma nel 1916 e riprodotto in Fig. 7. Poco tempo dopo lo attende una pessima sorpresa, conseguenza del fatto che l'Italia è ormai schierata in opposizione agli imperi centrali. Come documenta ampiamente Simona Salustri in una recente pubblicazione [32], contro di lui iniziano insinuazioni e sospetti che si tramutano presto in accuse esplicite di complicità con il nemico e di spionaggio in cambio di denaro, per cui già nel dicembre 1916 il prefetto dell'Umbria arriva a scrivere al Ministero che la presenza di Dessau e quella della sua famiglia non sono gradite in Italia. La situazione precipita poi nel corso del 1917 e, dopo la rotta di Caporetto, un gruppo di studenti scrive al Rettore una lettera, riprodotta e trascritta in [1], p.59, chiedendone la rimozione a causa della sua origine tedesca. Malgrado il tentativo di resistere di Dessau, che può ben affermare di essersi sempre comportato in modo leale verso la sua nuova patria di cui è cittadino da oltre vent'anni, il Rettore dell'Università di Perugia lo sospende dal servizio e gli viene applicata una forte riduzione dello stipendio. La mancanza di risorse economiche ed il clima ostile dovuto alle vicende belliche fa sì che la famiglia Dessau decida di trasferirsi a Firenze, affidandosi all'aiuto di amici del posto. Questo esilio forzato durerà per oltre due anni e sarà segnato da forti ristrettezze economiche, tanto che Bernardo sarà costretto a cercare lavoretti di fortuna e perfino a vendere un prezioso crogiolo di platino che suo padre gli aveva lasciato in eredità. Nel 1918 egli scrive un'accorata lettera al suo punto di riferimento di sempre, Augusto Righi, per chiedergli un aiuto a superare il momento difficile e trovare, se possibile, un lavoro [7]. Finalmente, dopo un'opportuna inchiesta condotta dalla commissione accademica, anche con accertamenti bancari per verificare eventuali finanziamenti esteri, al termine del 1919 Dessau viene reintegrato nell'insegnamento presso l'Università di Perugia e nella direzione del Gabinetto di Fisica, perciò scrive di nuovo a Righi per ringraziarlo del sostegno morale negli anni che, confessa, sono stati i più difficili della sua vita. La famiglia rimarrà a Firenze ancora un altro anno, prima di raggiungerlo a Perugia, anche per dare modo ai figli di completare l'anno scolastico nella sede toscana.

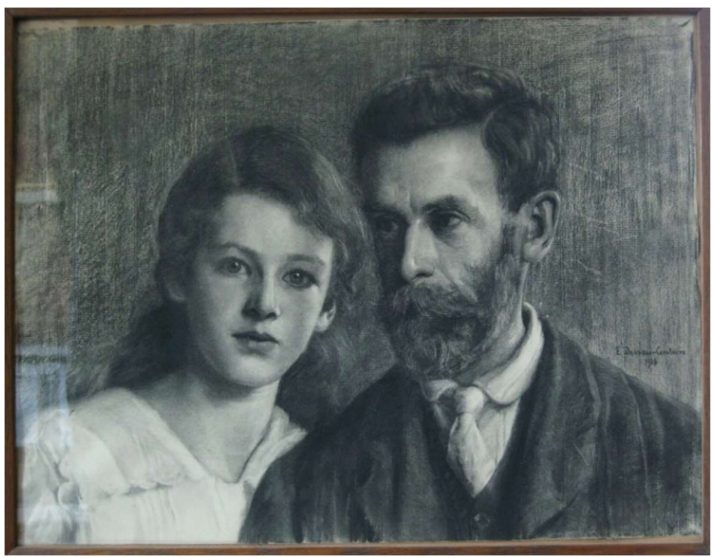

**Fig. 7**. Bernardo con la figlia Fanny, nel 1916. Ritratto della moglie Emma Goitein Dessau. Collezione Umberto Steindler.

**3.2 Il secondo quindicennio perugino, l'apice della carriera e le fugaci soddisfazioni accademiche.**

Dal 1920 inizia il secondo quindicennio perugino della famiglia Dessau. Essi si stabiliscono in affitto in un villino di Via Pompili, nel quartiere di Monteluce, prossimi al Policlinico e alla Facoltà di Medicina, i cui studenti sono suoi allievi per quanto riguarda la Fisica. Egli è ormai un professore ordinario ricco di esperienza e viene anche coinvolto in compiti di responsabilità, come la direzione dell'osservatorio meteorologico, la direzione della scuola di Farmacia o la partecipazione a commissioni di concorso per il reclutamento di nuovo personale. In Fig. 8 è riprodotta una foto di metà anni venti, dove si riconosce Dessau con un gruppo di colleghi della Facoltà di Medicina dell'epoca.

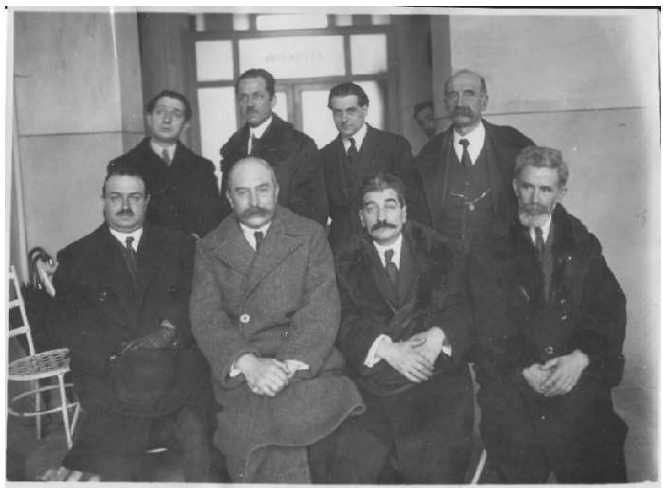

**Fig. 8** Gruppo di professori della Facoltà di Medicina dell'Università di Perugia alla metà degli anni venti del novecento. Fila in alto, partendo da sinistra: Lo Cascio, Bizzozzero, Aymerich, Agostini. Fila in basso, sempre da sinistra: Righetti, Polimanti, Filippi e Dessau. Archivio Storico Università di Perugia. Su concessione dell'Università degli studi di Perugia, soggetta a divieto di ulteriore riproduzione o duplicazione con qualsiasi mezzo.

Soprattutto in quegli anni, Dessau si costruisce la fama di grande didatta e innovatore, come ricorderà nel 1970 il prof. Franco Levi, suo successore sulla cattedra di Fisica di Perugia: *"Intere generazioni di professionisti, già studenti a Perugia di Medicina, Veterinaria, Farmacia e Scienze Agrarie ricordano l'insegnamento preciso e chiaro di Dessau: ricordano che ogni lunedì, mercoledì e venerdì, alle nove in punto, c'era lezione di Fisica… e ricordano un apparecchio che proiettava sullo schermo ciò che l'insegnante scriveva, seduto al proprio tavolo. Questo apparecchio è entrato da pochi anni nell'uso comune, ma Dessau l'aveva progettato e fatto costruire nel laboratorio dell'Istituto."* [33] Un altro aspetto interessante di questo periodo, che sottolinea in un certo senso come la personalità di Bernardo precorra i tempi, è il suo impegno come divulgatore scientifico al difuori dell'ambiente strettamente accademico. Già durante il suo ultimo anno di permanenza a Bologna aveva tenuto un corso di introduzione alla meteorologia presso l'Università popolare Giuseppe Garibaldi, (Ref. [1] p.12) una delle prime fondate in Italia a cura della

Società Operaia bolognese, per promuovere una coscienza nuova tra i lavoratori. Poi a Perugia nell'anno scolastico 1922/23 egli accetta un incarico per insegnare fisica e matematica agli studenti del Liceo Classico Mariotti ([1] p. 26 ). Inoltre, nel 1923 propone al Rettore, che lo autorizza volentieri, di poter svolgere un corso di introduzione all'elettrotecnica aperto al pubblico di cittadini interessati (la relativa documentazione, custodita dall'archivio storico dell'Università di Perugia, è riprodotta in [1]). Si rende anche disponibile per partecipare alle commissioni degli esami di maturità, come ad esempio nel 1924, quando viene nominato presidente di commissione in un liceo triestino, lasciando una indelebile impronta su alcuni studenti come Aurelia Gruber Benco che sarà poi giornalista e scrittrice [34].

Si tratta dunque di anni relativamente tranquilli e segnati dal pieno inserimento di Bernardo ed Emma nell'ambiente perugino, dove pian piano approfondiscono amicizie e legami. Essi si staccano dall'impegno sionista e si aprono al dialogo con esponenti di altre correnti religiose e filosofiche. Come annoterà la figlia Fanny: *"Decenni avanti il sorgere della «Amicizia Ebraico-Cristiana» essi avevano accolto e riunito attorno a se in reciproco rispetto i « giusti delle altre Nazioni » e mentre invitavano alla loro mensa sabbatica e alia celebrazione del Seder gli ospiti ebrei, essi aprivano le loro porte ed erano stretti in amicizia ai cattolici osservanti e frequentavano alti prelati e dignitari della Chiesa"* [18]. Questa rete di amicizie con personaggi del mondo intellettuale e religioso perugino, come il conte Zopiro Montesperelli, l'avvocato Wladimiro Babucci, il parroco di Monteluce don Ugo Palmerini, l'assistente della FUCI Mons. Luigi Piastrelli ([8]), poi anche i più giovani intellettuali antifascisti Aldo Capitini (filosofo pacifista e ispiratore della marcia Perugia-Assisi) e Walter Binni (letterato, socialista e deputato alla costituente), si rivelerà anche una sorta di rete di protezione negli anni tristi delle persecuzioni antisemite. Può sembrare sorprendente trovare esponenti del clero perugino nel gruppo dei suoi amici, così come pensatori liberi come Capitini, soprattutto se si pensa che la città e l'ateneo di Perugia sono stati caratterizzati da un forte sentimento anticlericale e da una pervasiva e ben organizzata presenza massonica, soprattutto nella classe dirigente. Ebbene, anche in questo Dessau si dimostra una personalità eccentrica e innovatrice, difficilmente incasellabile nel cliché dello scienziato e accademico di lungo corso, tipico della sua epoca. Questa apertura di vedute e capacità di spaziare sull'intero orizzonte della conoscenza e della cultura, contraddistingue anche una lectio magistralis che egli viene chiamato a tenere nel 1927, in occasione delle celebrazioni per il centenario di Alessandro Volta, durante la sedicesima riunione della Società italiana per il Progresso delle Scienze che si svolge proprio a Perugia, tra fine ottobre e i primi di novembre del 1927 ([35], parzialmente riportato in [1] p.26). In questo discorso mi sembrano rilevanti almeno due aspetti. Anzitutto, la profonda e brillante riflessione di apertura sul confronto tra la creazione di un artista, che è veramente "sua" in quanto riflette la soggettività e la creatività dell'autore, e la scoperta scientifica, che invece non appartiene allo scienziato scopritore, perché presto o tardi, fatalmente, sarebbe trovata per opera di altri. Come non leggere in questa riflessione la dinamica e la profondità della sana dialettica interna alla famiglia Dessau, dove Emma rappresenta l'artista che crea quadri, dipinti, brani musicali, mentre Bernardo si dedica con rigore e tutt'altro metodo all'indagine scientifica? E l'altro aspetto rilevante di questo discorso è la capacità di sintetizzare le acquisizioni sperimentali di Volta con un linguaggio forbito nella forma, ma facilmente accessibile anche ai non specialisti, pur essendo ricco e approfondito nei contenuti.

---

([8]) È significativa soprattutto l'amicizia con mons. Luigi Piastrelli, noto esponente del mondo cattolico che fu inizialmente vicino al movimento modernista, ma dal quale si distaccò dopo i contrasti con la gerarchia. Nel 1919 fu tra i fondatori del partito popolare, poi fu assistente del gruppo perugino della federazione universitaria cattolica italiana (FUCI) e assunse addirittura il ruolo di assistente nazionale della FUCI dal 1922 al 1925. Piastrelli frequentò casa Dessau già prima degli anni venti, come istitutore della figlia Fanny per le materie classiche, rimanendo poi a lungo amico di famiglia. Anche l'avvocato Wladimiro Babucci, amico e confidente dei Dessau, fu un intellettuale cattolico e antifascista di prima grandezza, anche se poi lasciò Perugia per Firenze. In questi rapporti di amicizia si palesa come, precorrendo i tempi del dialogo tra ebrei e cristiani, per Dessau fosse ben più importante la comune radice biblica e la fede nel Dio di Abramo, Isacco e Giacobbe, che non la diffidenza e i pregiudizi negativi stratificati durante i secoli.

A partire dal 1929, Dessau viene chiamato a far parte del CNR, presieduto da un paio d'anni da una sua vecchia conoscenza, quel Guglielmo Marconi che ha ormai scalato tutti i gradini del successo ed è uomo di punta del regime e della scienza nazionale. Dessau è poi addirittura invitato a Roma il 12 febbraio del 1931 alla solenne inaugurazione dei nuovissimi impianti di Radio Vaticana (come documentato in [1] p. 27), allestiti sotto la regia dello stesso Marconi, alla presenza di Benito Mussolini e di Papa Pio XI. Ci piace pensare che nel coinvolgere Dessau in queste avventure, Marconi abbia forse inteso, più o meno coscientemente, saldare un vecchio debito di riconoscenza, risalente a oltre trent'anni prima, con il duo Dessau – Righi (quest'ultimo morto nel 1920). In quello stesso periodo, Bernardo riceve anche il prestigioso incarico di redigere alcune voci per l'enciclopedia Treccani, nei volumi editi nel 1930/31 [36].

Pur essendo ovviamente alieno rispetto all'imperante ideologia fascista e condividendo tale sentimento con il gruppo di amici intellettuali perugini, non risultano dichiarazioni o atti che attestino pubbliche prese di distanza. Anzi, nel fascicolo personale presso l'Università di Perugia si trovano le copie dei verbali dei giuramenti di fedeltà alla corona d'Italia del 1927 e poi ancora quello del 1931 che include anche la fedeltà al regime fascista, come mostrato in Fig. 9, che egli rende al pari della quasi totalità dei docenti universitari italiani. L'unico docente perugino che si rifiutò di giurare, pagando il prezzo dell'espulsione dall'Università, fu il giurista Edoardo Ruffini, come ben ricostruito in [37], p. 234.

Nel 1935, ormai settantaduenne, è tempo di essere posto in congedo per raggiunti limiti di età. Non prima, però, di veder pubblicata la seconda edizione del terzo volume del Manuale di Fisica, dove egli inserisce le principali acquisizioni della nascente fisica quantistica. In Europa spira ormai un impetuoso vento antisemita ed in Germania le discriminazioni sono già entrate estensivamente nel corpus legislativo. L'Italia si incammina sullo stesso percorso e nel 1937 viene purtroppo improvvisamente a mancare Marconi, che avrebbe potuto forse fare qualcosa, data la sua vicinanza al mondo anglosassone, per sconsigliare a Mussolini l'abbraccio mortale con Hitler. Prima che la situazione precipiti, Dessau fa in tempo, comunque, a provare due grandi soddisfazioni. Anzitutto, la nomina a Professore Emerito, conferita dal Ministro dell'Educazione Nazionale su richiesta del Consiglio della Facoltà di Medicina del 5 luglio 1936, sebbene dopo una procedura un po' rallentata. Infatti, come si evince dalle copie delle lettere presenti ancora nel fascicolo personale presso l'Archivio universitario (alcune riprodotte in [1]), accade che il nuovo Rettore Paolo Orano, noto esponente di spicco del regime fascista ([9]), trasmette al Ministero nel luglio del 1936 la

---

([9]) Paolo Orano, deputato fascista di lungo corso, fu uno dei fondatori della Scuola fascista di giornalismo presso la Facoltà Fascista di Scienze Politiche, dove insegnava "Storia e dottrina generale del fascismo", e di cui divenne preside nel 1933, per poi assumere la carica di Rettore nel 1935. Ebbe richiesta del Consiglio di Facoltà di Medicina dimenticandosi (riteniamo lecito supporre volontariamente) di esprimere il suo parere. Accade quindi che il Ministero, in data 30 dicembre, scriva al Rettore ricordandogli che è necessario che egli esprima il suo giudizio. Non avendo pronta risposta, il 20 gennaio il Ministero sollecita di nuovo il Rettore a esprimersi ed ecco che, finalmente, il 23 gennaio Orano comunica il suo parere favorevole. A quel punto il Ministro, il 12 febbraio, scrive al Rettore di comunicare all'interessato che, con Decreto Reale in corso, Bernardo Dessau è nominato professore emerito. Orano compie dunque l'ultimo passo del dovere di ufficio e termina la sua comunicazione "con distinti saluti fascisti". Malgrado questa chiusa possa essere suonata piuttosto beffarda a Dessau, egli risponde con una lettera di ringraziamento al Rettore, dichiarando che si tratta della *"più bella ricompensa che ha potuto trovare la mia lunga attività di insegnante presso la nostra Università"*. In quello stesso periodo egli riceve pure l'onorificenza di Cavaliere dell'Ordine della Corona d'Italia [1] e la sua soddisfazione raggiunge il massimo.

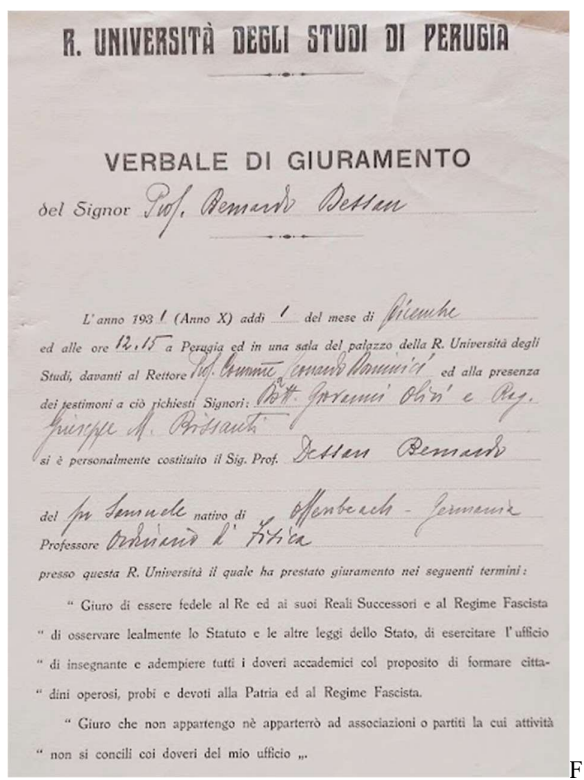

**Fig. 9** Copia del verbale del giuramento di fedeltà alla corona ed al regime fascista reso da Dessau nel 1931. Archivio dell'Università di Perugia, Personale docente cessato, 1-1, fasc. 3, anno 1935. Su concessione dell'Università degli studi di Perugia, soggetta a divieto di ulteriore riproduzione o duplicazione con qualsiasi mezzo.

---

un ruolo di primo piano nella diffusione di teorie antisemite in Italia. Il suo deciso progetto di fascistizzazione dell'Ateneo perugino è descritto nelle Ref. [32], p. 125 e [37], p 240.

Ma la gioia di questi riconoscimenti è fortemente smorzata dal clima che si sta facendo davvero pesante per gli ebrei, sia in Germania che in Italia, e già ai primi di febbraio del 1938 la moglie Emma scrive all'amico di famiglia avvocato Wladimiro Babucci che ormai risiede a Firenze: *"Noi sentiamo tanto, forse troppo, le sofferenze dei nostri correligionari… È tornato un medio evo per loro con torture più raffinate, rivestite di belle parole. Siamo in un periodo di regresso ed anche questo è opprimente."* [38].

### 4. Le leggi razziali, la guerra e il triste epilogo (1938-1949)

Dall'estate del 1938, com'è ben noto, si abbatte con violenza anche in Italia l'ombra oscura delle leggi razziali e gli ebrei sono espulsi dalle accademie e dalle società scientifiche. Per quanto riguarda la SIF, dal censimento di auto-denuncia compiuto nel settembre del 1938, quando vengono ricevute indietro 163 schede sulle 209 spedite ai soci, vengono individuati complessivamente 15 soci da radiare, tra cui appunto Bernardo Dessau e Giorgio Todesco che è il suo primo successore nella cattedra di fisica di Perugia [39]. Se ne aggiungono poi altri quattro, appartenenti al gruppo che non aveva risposto al primo giro. Pochi anni prima, nel 1934, Dessau era stato del resto costretto a lasciare la Deutschen Physikalischen Gesellschaft (DPG) Società Tedesca di Fisica, di cui era stato anche membro fin dal 1905 ([10]), visto che le leggi razziali in Germania dispiegarono i loro effetti già tra il 1934 e il 1935, anno delle leggi di Norimberga. Un'atra società scientifica italiana da cui Dessau risulta cacciato nel 1938 (al pari di suo figlio Gabor, geologo), in applicazione delle leggi per la difesa della razza, è la Società Italiana per il Progresso delle Scienze, a cui aveva partecipato attivamente fin dal 1907 [39]. Presso l'Università di Perugia, il Rettore Paolo Orano si dimostra ovviamente molto sollecito ad applicare le disposizioni razziali, quindi i tre professori ebrei in servizio vengono espulsi con effetto immediato ed anche Dessau, ormai emerito, è cancellato dal corpo accademico ed escluso dalla vita universitaria [32, 37]. Un penoso episodio, rappresentativo di tante altre vessazioni dell'epoca, è quello relativo all'apparecchio radio che viene sequestrato e sigillato, in casa Dessau, nel 1941. Come evidenziato dal documento, conservato nell'Archivio di Stato di Perugia, pubblicato e trascritto in [1] p.29, Bernardo scrive subito dopo al questore una tristissima lettera, chiedendo che gli sia consentito almeno di vendere tale apparecchio, ormai non più utilizzabile, all'amico Zopiro Montesperelli, a causa delle condizioni di forte indigenza della famiglia Dessau. Sembra davvero un paradosso che colui che aveva contribuito, in gioventù, a chiarire i meccanismi fisici alla base delle radiocomunicazioni, venga crudelmente privato dell'apparecchio radio che permette ai due anziani coniugi di godere di un po' di musica (essendo anche stati espulsi per motivi razziali dalla gloriosa Accademia dei Filedoni, che avevano a lungo frequentato a Perugia per coltivare la passione musicale) e di rimanere in contatto con il mondo. Un altro colpo duro per Dessau è quello dell'inclusione del suo nome, nel 1942, nella lista degli "autori non graditi" [40] per cui anche i volumi del suo Manuale di Fisica, che tanto lavoro e tanta cura gli sono costati, vengono ritirati dal commercio e considerati libri proibiti. Malgrado ciò, in questi ultimi anni di vita, segnati dalla solitudine e dalla malattia, Bernardo non rinuncia comunque alla sua attività intellettuale ed al suo interesse per la fisica. Lavora quotidianamente alla redazione di una nuova opera ([11]), rimasta purtroppo inedita ed andata smarrita, sulla quale offre la sua testimonianza il prof Franco Levi, che ne parla nel discorso di commemorazione tenuto presso l'Università di Perugia in occasione dei venti anni dalla scomparsa di Dessau [33]: *"L'attività di Dessau non si esaurì con la fine del suo insegnamento: nell'ultimo decennio della sua vita, nel difficile periodo della seconda guerra mondiale, rimasto solo con i propri libri e i propri pensieri, tormentato da gravi sofferenze fisiche e dall'angoscia causata dalla persecuzione razziale, si diede a scrivere una nuova opera "Il divenire di una scienza: la Fisica". Questo libro è rimasto a tutt'oggi inedito: siamo tra i pochi che hanno avuto la possibilità di consultarne il manoscritto, che è completo e per la massima parte corretto dall'Autore stesso….. Esso ci offre un panorama dinamico dell'evoluzione dei concetti fondamentali della Fisica, dai primordi fino all'epoca della sua compilazione… Si tratta di un testamento spirituale, di un avvertimento rivolto ai giovani…".* Purtroppo, il manoscritto è andato smarrito dopo la morte di Dessau e solo un capitolo, riguardante il Calorico, è stato pubblicato postumo [25].

---

([10]) La scomparsa del nome di Dessau dall'elenco dei membri ufficiale della Società, dal 1934 in poi, si evince dalla consultazione delle liste di membri, accessibile dal sito web della DFG. https://www.dpg-physik.de/ueber-uns/profil-und-selbstverstaendnis/archiv-der-dpg

Nonostante le crescenti restrizioni e persecuzioni di quegli ultimi anni di vita, Bernardo ed Emma rimangono nel villino affittato in Via Pompili fino all'estate del 1943, dopodiché la situazione precipita e sono costretti a nascondersi per scampare ai rastrellamenti nazifascisti. I figli per fortuna sono all'estero, Fanny è a Tel Aviv, dove si è sposata ed è ormai mamma di due bambini, e Gabor in India prigioniero degli inglesi, al riparo dalla tempesta che squassa l'intera Europa. Bernardo viene nascosto nella clinica medica del policlinico universitario, grazie al direttore, amico e collega, prof. Fedele Fedeli ([1], p.31). Emma si rifugia invece in campagna, dove dei conoscenti la ospitano in un rifugio di fortuna ricavato in un annesso. Dopo quarantuno anni di vita comune, i due coniugi non solo sono separati per parecchi mesi, ma vivono anche nell'angoscia di non avere notizie l'uno dell'altro, se non per qualche rara ambasciata di amici fidati, come Binni e Capitini [18]. Finalmente, entrambi escono dalla clandestinità dopo la liberazione di Perugia, avvenuta il 20 giugno del 1944 ([12]). Un increscioso episodio accade nel successivo mese di dicembre, come ben ricostruito in [37] p. 245, quando Dessau è costretto a scrivere una risentita lettera al nuovo Rettore, il democristiano Giuseppe Ermini che avrebbe governato per un trentennio, per lamentare il suo mancato invito alla cerimonia di inaugurazione del nuovo anno accademico, il primo dopo la liberazione. Afferma Dessau: *"voglio ancora credere ad una dimenticanza – profondamente dolorosa per chi, come me, ha dovuto sopportare assieme ai propri cari le inique leggi*

---

([11]) In una lettera dei primi anni di guerra, indirizzata alla figlia Fanny che risiede a Tel Aviv con la sua nuova famiglia, in possesso del nepote Umberto Steindler, dopo aver lamentato i suoi problemi di salute, Bernardo annota: *"quasi quotidianamente continuo a scrivere una mezza paginetta di quel libro che non sarà mai terminato, ma intanto vi convincerà che ho ancora un po' di forza. Coraggio dunque e andiamo avanti."* Vale la pena notare che parte di questa lettera, dove verosimilmente si fa cenno alle difficoltà del momento storico, è oscurata con inchiostro nero dalla censura fascista.

([12]) È interessante notare come nel fascicolo personale di Dessau presso l'Archivio storico dell'Università di Perugia, ricco di tantissimi documenti relativi anche a questioni del tutto marginali, come le brevi malattie, le trasferte, gli spostamenti di una singola lezione, ecc., non c'è nulla che riguardi i provvedimenti discriminatori e le espulsioni di cui egli fu oggetto a partire dal 1938. Del resto, è ben noto che dopo la guerra ci fu una sorta di tacito accordo, che coinvolse sia i carnefici che le vittime, per favorire una sorta di rimozione collettiva a riguardo degli orrori vissuti. Colpisce, ad esempio, che anche in interventi commemorativi come quello per i vent'anni dalla morte di Dessau, riportato in Ref. [33], a opera del prof Franco Levi, successore di Dessau e anch'egli ebreo, ai tristi eventi del periodo successivo al 1938 si faccia appena un fugacissimo cenno.

*razziali, sempre fiducioso nel ritorno della libertà e [della] giustizia – piuttosto che di ritenere possibile, proprio nell'ora del rinnovamento, un tentativo di mantenere, sulla mia persona, le menomazioni legali oramai abolite".* Ermini si precipita a rispondere, scusandosi sinceramente e assicurando che si è trattato in effetti di semplice dimenticanza, da non interpretare assolutamente in altro modo.

Ma ormai l'età avanzata ed i terribili eventi attraversati, a cui si aggiungono le strazianti notizie dei tanti familiari e parenti assassinati nei campi di sterminio nazisti, comportano conseguenze irreversibili. Bernardo è in uno stato di salute molto precario e a fatica conserva un po' di autonomia per badare a sé stesso, tornando nell'abitazione di via Pompili. Emma è provata psicologicamente e viene ricoverata nella clinica psichiatrica di Perugia. Seguono anni di grande silenzio e solitudine per entrambi. Bernardo può contare soltanto sull'aiuto della signora Bonucci, che abita al piano inferiore della sua stessa palazzina, oltre che sulle visite mensili della figlia Fanny che è rientrata da Tel Aviv e abita ormai stabilmente a Genova. Lamenta che in quattro anni, soltanto un paio di volte è stato accompagnato a visitare sua moglie in ospedale. Muore il 17 novembre 1949, all'età di ottantasei anni, e sulla lapide della tomba vuole che, accanto al suo nome, venga scritto: "fisico, scienziato, maestro". Emma gli sopravvive per quasi un ventennio, rimanendo confinata nella clinica psichiatrica perugina, visitata di tanto in tanto da figli e nepoti, e termina la sua esistenza il 17 settembre 1968 ([13]). La sua tomba è accanto a quella del marito, nella sezione ebraica del cimitero monumentale di Perugia.

Negli anni cinquanta, presso l'Università di Perugia è stata dedicata a Bernardo Dessau l'aula nobile del Rettorato (su iniziativa dell'ex-collega Lo Cascio, come ricostruito in dettaglio in [1] p. 35), dov'è tuttora collocata la lapide che lo ricorda così: SERENO E GRANDE SPIRITO DI MAESTRO CHE OGNI DOLORE DELLA TORMENTATA VITA SUPERÒ NEL FERVORE DELLA SUA MISSIONE PERCHÉ COL SUO NOME IN QUESTE AULE DI LUI RIMANGA L'ESEMPIO E IL RIMPIANTO.

Nel settembre 2022, in occasione del XLII Congresso Nazionale della Società Italiana degli Storici della Fisica e dell'Astronomia (SISFA), presso il Dipartimento di Fisica e Geologia di Perugia è stata allestita una esposizione degli strumenti antichi del Gabinetto di Fisica [41], molti dei quali utilizzati, acquistati o restaurati proprio da Bernardo Dessau, in mezzo ai quali fa bella mostra di sé il pannello a lui dedicato, come illustrato in Fig. 10 ([14]).

---

([13]) Dai racconti di Umberto Steindler sulle visite alla nonna Emma e dalla lucidità ed eleganza di alcune lettere scritte da Emma durante il periodo di permanenza nella clinica psichiatrica, risulta davvero difficile comprendere le ragioni di un internamento così lungo, senza alcuna via d'uscita se non la morte.

([14]) In occasione del congresso SISFA del settembre 2022 è stata anche messa in scena la prima rappresentazione di un bellissimo e intenso documentario teatrale, su Bernardo Dessau, scritto e diretto da Paola Tortora di Vintulera Teatro e interpretato dal collega fisico Alessio Stollo. Questo spettacolo è stato anche proposto, in occasione della Giornata della Memoria 2023, a una platea di circa cinquecento studenti dei licei perugini, riscuotendo unanime apprezzamento, ed è intenzione degli autori proporlo in futuro anche in altre sedi o città italiane per far conoscere la vicenda umana e scientifica di Dessau.

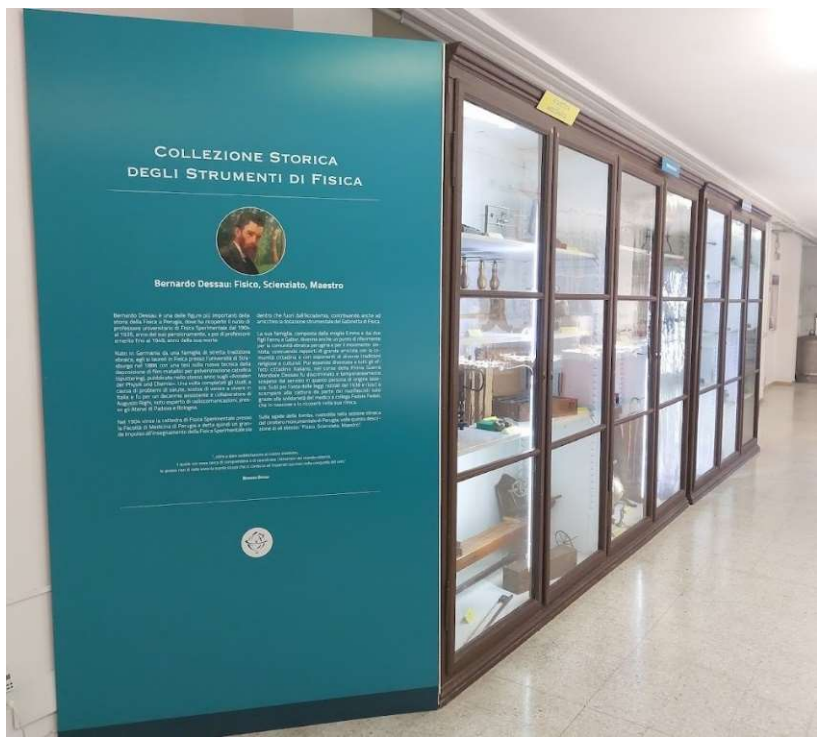

**Fig. 10** Fotografia dell'atrio del Dipartimento di Fisica e Geologia dell'Università di Perugia, dove nel settembre 2022 è stato collocato un pannello dedicato alla vicenda biografica e scientifica di Bernardo Dessau, accanto alla rinnovata esposizione degli strumenti antichi del Gabinetto di Fisica.

### 5. Conclusioni

A distanza di settantacinque anni dalla morte di Bernardo Dessau, abbiamo voluto contribuire, con queste pagine, a sollevare il velo di oblio che ha reso di fatto sconosciuta la sua figura per larga parte della comunità dei fisici italiani. Un'attenzione particolare è stata dedicata al periodo bolognese, dal momento che esso è stato finora la parte più ignorata del percorso biografico e scientifico di Dessau, essendo relativa alla sua fase giovanile, per di più vissuta all'ombra di giganti del calibro di Augusto Righi e Guglielmo Marconi. Abbiamo inteso mettere in evidenza le sue qualità di ricercatore, di didatta e di divulgatore nel campo della fisica, ma anche quelle di fine intellettuale con interessi in altri campi del sapere, di costruttore di ponti tra identità diverse, di uomo giusto ancorato alle tradizioni familiari, ma capace di nuove sintesi. In diverse fasi della sua vita ha pagato prezzi alti per rimanere fedele alle sue multiple appartenenze e, limitandoci al campo della scienza, non possiamo esimerci dal notare come le potenzialità di quel giovane forgiato da maestri di prima grandezza della fisica mitteleuropea, come Helmholtz e Kundt, arrivato in Italia da Strasburgo a fine ottocento, non sembrano essersi dispiegate pienamente durante la sua vita accademica. Probabilmente ciò è dipeso da caratteristiche della sua personalità che mal si conciliavano con quelle richieste, all'epoca, per una brillante carriera nella comunità della fisica italiana. Si può quindi condividere nella sostanza quanto aveva già annotato, mezzo secolo fa, Alberigi Quaranta sul Giornale di Fisica: "*Animato da buone intenzioni e desideroso di proseguire il suo lavoro scientifico, egli non aveva quel tipo di inserimento nella società italiana, quell'energia e aggressività necessarie ad agire efficacemente nell'ambito della ricerca e dell'insegnamento della Fisica in Italia. Questa situazione fece sì che una persona indubbiamente capace non venisse affatto utilizzata, bensì mantenuta in uno stato di isolamento pressoché completo*". [42]

Durante il cinquantennio della sua attività scientifica Bernardo Dessau ebbe la fortuna di assistere da vicino alle strabilianti scoperte e novità sul fronte della fisica relativistica, della meccanica quantistica e delle telecomunicazioni, e ne restò affascinato. Ma sperimentò anche la mala sorte di essere scosso, a più riprese, dai terribili eventi storici e sociali che hanno caratterizzato la prima metà del novecento. Ne uscì piegato, ma non vinto, e conservò fino alla fine nobiltà d'animo, finezza di spirito e, soprattutto, curiosità intellettuale per come si stava evolvendo la descrizione del mondo fisico. E anche se, purtroppo, è andata perduta l'ultima sua opera dedicata a narrare questa evoluzione, rimane integra ed esemplare la sua testimonianza di fisico, scienziato, maestro.

## 6.    Ringraziamenti